\documentclass[numberedappendix]{emulateapj}

\usepackage{graphicx}
\usepackage{natbib}
\usepackage{longtable}

\shorttitle{Evolution of BCGs structural parameters}
\shortauthors{Ascaso et al.}

\begin{document}

\title{Evolution of BCGs structural parameters  in the last $\sim$6 Gyr: feedback processes versus merger events.} % from local to high redshift

\author{B. Ascaso$^{1}$}
\affil{$^{1}$ Department of Physics, University of California, Davis, One Shields Avenue, Davis, CA 95616, USA}
%\and
\author{J. A. L. Aguerri$^{2,3}$,  J. Varela,  A. Cava$^{2,3}$}
\affil{$^{2}$ Instituto de Astrof\'isica de Canarias, C/ V\'ia L\'actea s/n, 38200 La Laguna, Tenerife, Spain\\
$^{3}$ Departamento de Astrof\'isica, Universidad de La Laguna E-38205, La Laguna, Tenerife, Spain}
%\and
\author{D. Bettoni$^{4}$}
\affil{$^{4}$ INAF- Osservatorio Astronomico di Padova, Vicolo Osservatorio 5, 35122, Padova, Italy}
%\and
\author{M. Moles$^{5,6}$}
\affil{$^{5}$ Instituto de Astrof\'isica de Andaluc\'ia-CSIC, Glorieta de la Astronom\'ia s/n, 18008, Granada, Spain\\
$^{6}$ Centro de Estudios de F\'isica del Cosmos de Arag\'on(CEFCA), C/General Pizarro 1, 44001,Teruel, Spain}
\author{M. D'Onofrio$^{7}$}
\affil{$^{7}$ Dip. Astronomia, Universita di Padova, Vicolo Osservatorio 2, 35122 Padova, Italy}

\begin{abstract}

We present results on the evolution in the last 6 Gyr of the structural parameters of two samples of brightest cluster galaxies (BCGs). The nearby sample of BCGs consist on 69 galaxies from the WINGS survey spanning a redshift range of 0.04$<$z$<$0.07. The intermediate redshift (0.3$<$z$<$0.6) sample is formed by 20 BCGs extracted from the Hubble Space Telescope archive. Both samples have similar spatial resolution and their host clusters have similar X-ray luminosities. 
We report an increase in the size of the BCGs from intermediate to local redshift. However, we do not detect any variation in the S\'ersic shape parameter in both samples. These results are proved to be robust since the observed tendencies  are model independent.
We also obtain significant correlations between some of the BCGs parameters and the main properties of the host clusters. More luminous, larger and centrally located BCGs are located in more massive and dominant galaxy clusters. These facts indicate that the host galaxy cluster has played an important role in the formation of their BCGs.
We discuss the possible mechanisms that can explain the observed evolution of the structural parameters of the BCGs. We conclude that the main mechanisms that can explain the increase in size and the non-evolution in the S\'ersic shape parameter of the BCGs in the last 6 Gyr are feedback processes. This result disagrees with semi-analytical simulation results supporting that merging processes are the main responsible for the evolution of the BCGs until the present epoch.

\end{abstract}


\section{Introduction}

The Brightest Cluster Galaxies (BCGs) are the most luminous and massive stellar systems in the Universe. BCGs are usually found very close  to the center of the clusters of galaxies determined from X-ray or gravitational lensing observations \citep{jones84,smith05}. This suggests that the brightest cluster  members have settled down in the potential well of the cluster (but see also  \citealt{coziol09}). These special objects possess a number of singular properties, being their origin and evolution directly related with the mass assembly in galaxy clusters.

BCGs luminosities are remarkably homogenous, as noticed first by \cite{humason56}. A number of works \citep{sandage72a,gunn75,hoessel85,postman95}, verified their high luminosities and small scatter in absolute magnitude and consequently, proposed them as {\it standard candles} for measuring cosmological distances. In fact, they were originally used to increase the range of  Hubble's redshift - distance law \citep{sandage72a,sandage72}.

Furthermore, there are numerous pieces of evidence showing that BCGs are not extracted from the same luminosity distribution function as  normal galaxies \citep{tremaine77,loh06,ascaso08,ascaso08b,lin09}.  Those differences could be related with the formation of BCGs in a different way than normal elliptical galaxies. There are indications that the environment plays an important role in the formation of BCGs due to their special location in the cluster. Thus, several works found correlations between the BCGs luminosity and the mass or the X-ray luminosity of the clusters \citep{brough02,nelson02,lin04,whiley08,sanderson09}. \cite{lambas88} even discovered an alignment between the major axis of the BCGs and the distribution of galaxies around the clusters located in 15 Mpc scales.

On a different perspective, considerable observational evidence \citep{bower92a,aragon-salamanca93,stanford98,vandokkum98,miley06,vandokkum10} suggest that the stars of giant elliptical galaxies were formed at high redshift, and have been passively evolving to the present day, but see also \cite{mancini10}.  Nevertheless, this passive evolution is different from normal elliptical galaxies because stellar population studies show that BCGs are more metallic and have larger $\alpha$-enhancement than normal elliptical galaxies \citep{loubser09}. This passive evolution of the stellar population is in apparent contradiction with some studies showing an evolution of the size and mass of BCGs.  For instance, \cite{aragon-salamanca98} found that BCG galaxies have grown their masses in the last 8 Gyr.  \cite{nelson02} reported a growth of $\sim$ 2 at z$\sim$ 0.5 and \cite{bernardi09} showed that BCGs at z $\sim$ 0.25 are 70\% smaller in size than nearby ones.
 
On the other hand, the surface brightness profiles of BCGs are usually well fitted by de Vaucouleurs or S\'ersic profiles even at large radii \citep{graham96}, similar to normal elliptical galaxies \citep{trujillo01,graham03,aguerri04,kormendy09}. Nevertheless, some of them show an excess of light, usually called envelopes, over the r$^{1/4}$ profile at large radii \citep{matthews64,oemler73,oemler76,schombert86,schombert87,schombert88,gonzalez05,seigar07}. These envelopes show low surface brightness and large spatial extension \citep{zibetti05}. Although the origin of such extended envelopes is still not completely clear, \cite{patel06} claimed that the extended stellar haloes of the BCGs are likely from the BCGs themselves (see also the works on M87;  \citealt{arnaboldi04,doherty09}). These extended stellar haloes are not part of the so-called intracluster light (ICL) formed by non-bounded stars and observed in some nearby clusters \citep{arnaboldi02,aguerri05,gerhard07, castro-rodriguez09}. They are formed by stars gravitationally bounded to the BCG. Nevertheless, the origin of the extended envelopes could be related to the origin of the ICL (e.g. \citealt{murante07}).

Different theories have been proposed to explain the observational properties of BCGs and give a framework about their formation. BCGs were proposed to be formed by the accumulation of tidal stripped debris from clusters of galaxies \citep{ostriker75,mcglynn80,merritt85}. Galaxy cannibalism in the central regions of galaxy clusters can also produce massive galaxies similar to BCGs  \citep{ostriker77}. Also, \cite{fabian82} proposed gas cooling flows presented in the centers of galaxy clusters as the responsible for creating these systems. 

During the last decade the cold dark matter (CDM) scenario has been considered the most appropriate in order to explain the structure formation in the Universe. This galaxy formation scenario can also explain the formation of BCGs. Thus, \cite{dubinski98} showed that natural merging process of dark matter haloes in a hierarchical model can produce central galaxies with similar surface brightness and velocity dispersion as the observed ones. Recently, hierarchical simulations of structure formation have shown that the stellar component of BCGs was formed at early epochs (50\% at z$\sim$5 and 80\% at z$\sim$3) in separated galaxies which then, accreted material to form the BCG through dry mergers  \citep{deLucia07b}. This implies that most of the stars located actually in BCGs were not formed in situ. In contrast, they were accreted from galaxy satellites over the formation history of the galaxy. These accreted stars built up the extended haloes observed on BCGs \citep{abadi06,murante07}. Recently, it was found that the period of mass growth of BCGs is shorter than  the expected from numerical simulations \citep{collins09}.

In this paper, we have explored the properties of a sample of nearby BCGs from WINGS (WIde-field Nearby Galaxy-cluster Survey,  \citealt{fasano06}). We have analyzed their surface brightness distribution, and performed a study of their structural parameters and morphology. We have studied the evolution of all those properties by comparing them with a higher redshift BCGs sample (0.3$<$ z $<$ 0.6) imaged with the Advanced Camera for Surveys (ACS) at the Hubble Space Telescope (HST). Additionally, we have also compared the structural parameters that define the BCGs with the global parameters of the host clusters. This dataset allows to investigate the evolution of the structural parameters of BCGs in a period of $\sim$ 6 Gyr and give valuable indications about the mass assembly in galaxy clusters.

The structure of this paper is as follows. In Section 2, we present the BCGs samples we have analyzed in this paper. In Section 3, we analyze and explain the procedures used for fitting the galaxies surface brightness. In Section 4, we show the evolution with redshift of the BCGs structural parameters, magnitudes and envelope light. Section 5 is devoted to the search for relations between the BCGs and their host cluster properties. Finally, we show the discussion and conclusions of this work in Section 6 respectively. Throughout this paper we have adopted the same WINGS $\Lambda$CDM cosmology: H$_0$=70 km s$^{-1}$ Mpc$^{-1}$, $\Omega_m$=0.3 and $\Omega_{\Lambda}$=0.7.

\section[]{Data sample}

In this work, we have analyzed the population of BCGs  in two samples. On one hand, we have selected the BCGs in WINGS \citep{fasano06}. This cluster survey consist on 77 clusters in the redshift range of  0.04 $\leq z \leq$ 0.07, 36 of them were observed from the North hemisphere with the Wield Field Camera (WFC) mounted at the Isaac Newton Telescope (INT)-2.5m at La Palma, Spain, while the remaining 41 ones were imaged with the Wide Field Imager (WFI) in Max Planck Gesellschaft (MPG-ESO)-2.2m in La Silla, Chile. All clusters were imaged throughout the V band pass. The images were taken under seeing conditions of $\sim 1^{''}$, implying that the typical resolution for these images was $\sim$ 1 kpc. 

The WINGS clusters were selected from the X-ray ROSAT catalogues \citep{ebeling96} with X-ray fluxes $\ge 5.0\times 10^{-12} erg \, cm^{-2} s^{-1}$ in the 0.1-2.4 keV band and $|b| > 20$ deg. This survey has a compromise to obtain a large spatial coverage (around 1.6-2.7 Mpc radius) and depth (complete up to V$\sim$ 21.7 mag at 90\%, \citealt{varela09}). The analysis of the properties of such clusters can help to determine a zero point comparison in the properties of local clusters with respect to higher redshift surveys. The main properties of the BCGs in WINGS sample are listed in Table 1 in \cite{fasano10}.

We have excluded eight BCGs from this sample due to different issues. A193 has three galaxies interacting with the BCG. RXJ0058 and A2626 consist on a couple of galaxies with probably an AGN in one of them. In addition, the BCGs in A133, A160, A780, A3164 and IIZW108 are either too close to the edge of the chip or have closer stars. These facts make our fit not to converge to a good solution. Then, the final sample consists on 69 BCGs.

On the other hand, we have selected an intermediate redshift sample of 20 BCGs (0.3 $<$ z $<$ 0.6) extracted from the HST archive. These BCGs belong to a cluster sample observed with the Advanced Camera for Surveys (ACS) in the Hubble Space Telescope (HST) through the F814W band, spanning the same range of X-ray luminosity than WINGS. The high resolution of the ACS makes that the minimum scale that we can resolve in the intermediate redshift clusters is similar ($\sim$0.6 kpc) to the typical resolution in the WINGS sample. 

The ACS observations were carried out in Cycle 13 and 14 (proposals 10490 and 10152).  The BCGs were observed in single pointings of more than  2000 seconds. The original sample consisted of a complete, homogeneous sample of 72 X-ray clusters \citep{mullis03}. From this sample, 26 were observed with the snapshot program. We just selected the BCGs that were clearly the brightest from the cluster and had spectroscopic or photometric redshift available from the NED database. The final sample consists on 20 BCGs.

The galaxy clusters in the ACS sample were observed through a different band-pass than nearby clusters. In order to have the  same rest-frame magnitudes for nearby and intermediate redshift galaxies, we have transformed the F814W-band to V-band rest-frame using the following transformation:

\begin{equation}
V(0)-F814W(z)=(V-F814W)_{0} -k_{F814W} -EC_{F814W}
\end{equation}

where $k_{F814W}$ and $EC_{F814W}$ are the K-correction and evolutionary correction in the F814W band \citep{poggianti97}, and
$(V-F814W)_{0}$ is the rest-frame (V-F814W) color. The surface brightness of the galaxies was also corrected from cosmological dimming. 

\begin{figure}
\centering
\includegraphics[clip,angle=90,width=1.\hsize]{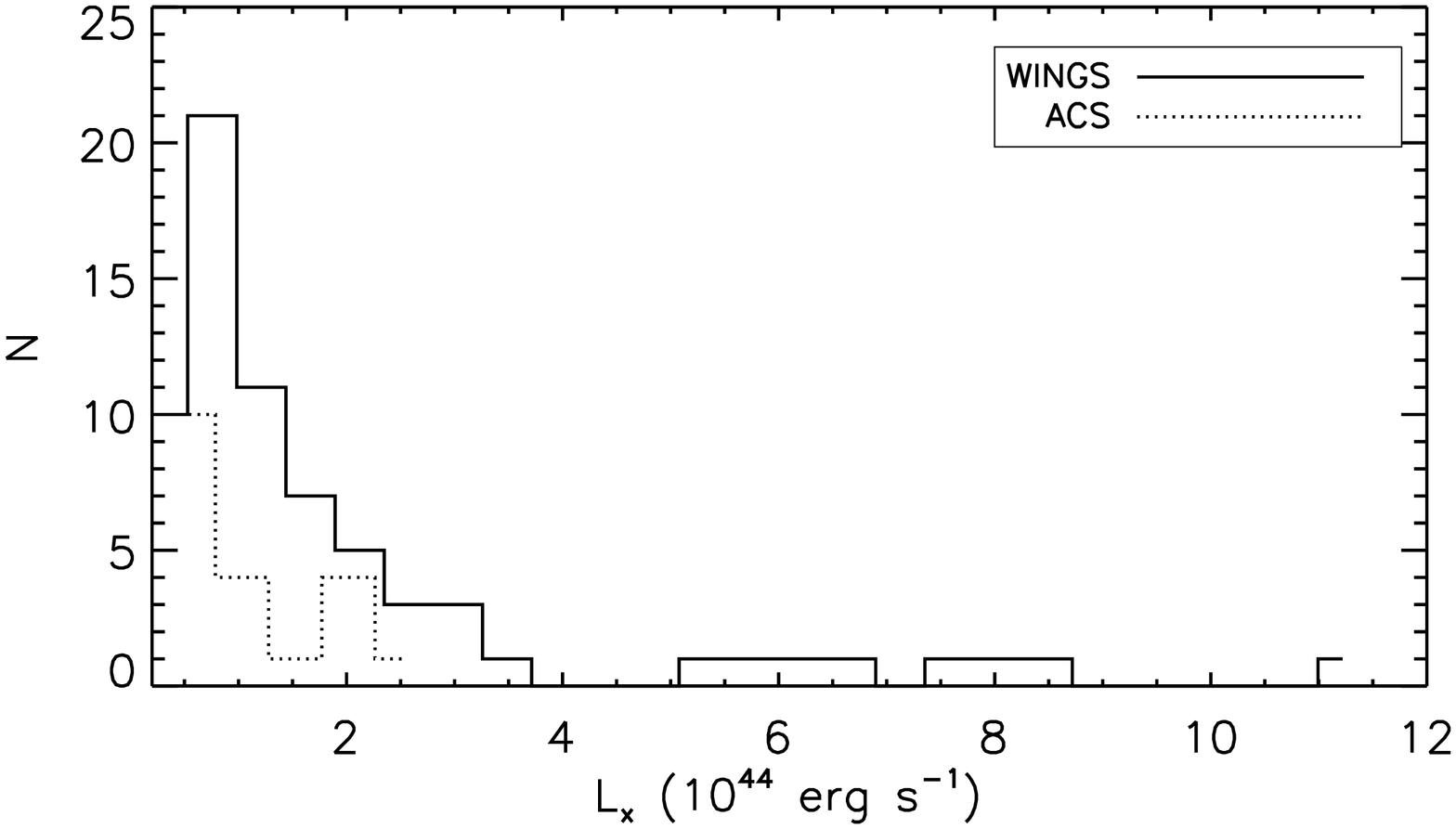}
\includegraphics[clip,angle=90,width=1.\hsize]{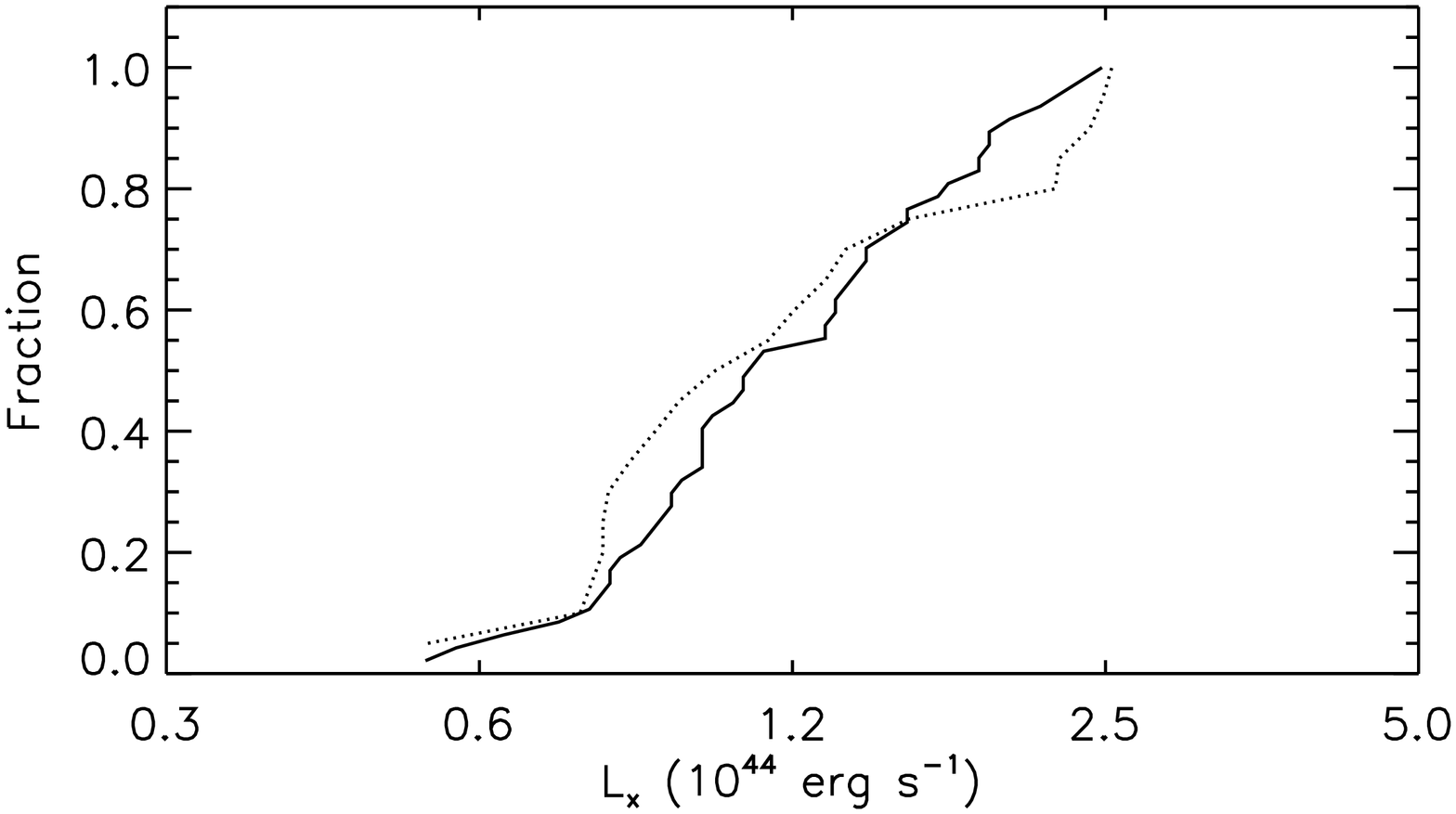}
\caption{$L_x$ distribution of the clusters sample in WINGS (solid line) and ACS (dotted line). The top panel shows the overall distribution for the whole ACS and WINGS sample, while the bottom panel refers to the cumulative distribution for the ACS sample and the WINGS X-ray luminosity restricted sample.}
\label{fig:lx}
\end{figure}

The top panel in Figure \ref{fig:lx} shows the X-ray luminosity distribution for the ACS clusters sample, with the X-ray  distribution function of the whole WINGS sample overplotted. The bottom panel in the same Figure refers to the cumulative function of the ACS clusters sample. We have also overplotted here the accumulated function of those WINGS clusters showing the same range of X-ray luminosity as the ACS sample, $5 \times 10^{43}<L_x<2.52 \times 10^{44} \,erg \,s^{-1}$. We have performed a Kolmogorov-Smirnoff (KS) test in these two samples, resulting that both distributions are  statistically similar. This implies that the global cluster properties (i.e. mass, velocity dispersion) are similar in the selected  nearby and intermediate redshift galaxy cluster samples. The cluster mass evolution has non effect in this selection since we have previously checked that there is not trend between $L_x$ of the host cluster and the BCG effective radius in both samples. In Tables  \ref{tab:WINGS} and \ref{tab:ACS}, we list the names of the BCGs in the WINGS and ACS sample respectively, together with their coordinates, X-ray luminosity and redshift of the host cluster.

\begin{table}
      \caption[]{WINGS BCGs sample}
     \[
        \begin{tabular}{lccccccl}
            \hline\noalign{\smallskip}
\multicolumn{1}{c}{\rm Name}&
\multicolumn{1}{c}{$\alpha$ (J2000)}&
\multicolumn{1}{c}{$\delta$ (J2000)}&
\multicolumn{1}{c}{$L_x$}&
\multicolumn{1}{c}{z}\\
\multicolumn{1}{c}{\rm }&
\multicolumn{1}{c}{hh:mm:ss}&
\multicolumn{1}{c}{dd:mm:ss}&
\multicolumn{1}{c}{10$^{44}$ erg/s}&
\multicolumn{1}{c}{}\\
\hline\noalign{\smallskip}           
{\rm A85} &      00:41:50.45  &  -09:18:11.5   &   4.28  &    0.0551 \\[-0.5ex]
{\rm A119} &     00:56:16.12  &  -01:15:19.0   &   1.65  &    0.0444 \\[-0.5ex]
{\rm A133} &     01:02:41.72  &  -21:52:55.4   &   1.82  &    0.0566 \\[-0.5ex]
{\rm A147} &     01:08:12.04  &  +02:11:38.2   &   0.28  &    0.0447 \\[-0.5ex]
{\rm A151} &     01:08:51.13  &  -15:24:23.0   &   0.52  &    0.0532 \\[-0.5ex]
{\rm A160} &     01:12:59.57  &  +15:29:28.8   &   0.19  &    0.0438 \\[-0.5ex]
{\rm A168} &     01:14:57.58  &  +00:25:51.1   &   0.56  &    0.0450 \\[-0.5ex]
{\rm A193} &     01:25:07.64  &  +08:41:57.2   &   0.79  &    0.0485 \\[-0.5ex]
{\rm A311} &     02:09:28.41  &  +19:46:36.2   &   0.41  &    0.0661 \\[-0.5ex]
{\rm A376} &     02:46:03.94  &  +36:54:19.1   &   0.71  &    0.0476 \\[-0.5ex]
{\rm A500} &     04:38:52.51  &  -22:06:39.0   &   0.72  &    0.0678 \\[-0.5ex]
{\rm A548b} &    05:45:29.62  &  -25:55:56.8   &   0.15  &    0.0416 \\[-0.5ex]
{\rm A602} &     07:53:26.61  &  +29:21:34.4   &   0.57  &    0.0619 \\[-0.5ex]
{\rm A671} &     08:28:31.66  &  +30:25:53.0   &   0.45  &    0.0507 \\[-0.5ex]
{\rm A754} &     09:08:32.39  &  -09:37:47.3   &   4.09  &    0.0547 \\[-0.5ex]
{\rm A780} &     09:18:05.68  &  -12:05:43.2   &   3.38  &    0.0539 \\[-0.5ex]
{\rm A957} &     10:13:38.27  &  -00:55:31.2   &   0.40  &    0.0451 \\[-0.5ex]
{\rm A970} &     10:17:25.71  &  -10:41:20.2   &   0.77  &    0.0591 \\[-0.5ex]
{\rm A1069} &    10:39:43.44  &  -08:41:12.3   &   0.48  &    0.0653 \\[-0.5ex]
{\rm A1291} &    11:32:23.22  &  +55:58:03.0   &   0.22  &    0.0509 \\[-0.5ex]
{\rm A1631a} &   12:53:18.41  &  -15:32:03.8   &   0.37  &    0.0461 \\[-0.5ex]
{\rm A1644} &    12:57:11.60  &  -17:24:34.0   &   0.04  &    0.0467 \\[-0.5ex]
{\rm A1668} &    13:03:46.60  &  +19:16:17.4   &   0.81  &    0.0634 \\[-0.5ex]
{\rm A1736} &    13:26:44.09  &  -27:26:21.8   &   1.21  &    0.0458 \\[-0.5ex]
{\rm A1795} &    13:48:52.51  &  +26:35:34.5   &   5.67  &    0.0633 \\[-0.5ex]
{\rm A1831} &    13:59:15.11  &  +27:58:34.5   &   0.97  &    0.0634 \\[-0.5ex]
{\rm A1983} &    14:52:55.33  &  +16:42:10.5   &   0.24  &    0.0447 \\[-0.5ex]
{\rm A1991} &    14:54:31.50  &  +18:38:32.8   &   0.69  &    0.0584 \\[-0.5ex]
{\rm A2107} &    15:39:38.92  &  +21:46:58.1   &   0.56  &    0.0410 \\[-0.5ex]
{\rm A2124} &    15:44:59.02  &  +36:06:33.9   &   0.69  &    0.0666 \\[-0.5ex]
{\rm A2149} &    16:01:28.11  &  +53:56:50.3   &   0.42  &    0.0679 \\[-0.5ex]
{\rm A2169} &    16:13:58.09  &  +49:11:22.3   &   0.23  &    0.0578 \\[-0.5ex]
{\rm A2256} &    17:04:27.22  &  +78:38:25.4   &   3.60  &    0.0581 \\[-0.5ex]
{\rm A2271} &    17:18:16.66  &  +78:01:06.2   &   0.32  &    0.0576 \\[-0.5ex]
{\rm A2382} &    21:51:55.62  &  -15:42:21.2   &   0.46  &    0.0641 \\[-0.5ex]
{\rm A2399} &    21:57:01.72  &  -07:50:22.0   &   0.51  &    0.0578 \\[-0.5ex]
{\rm A2415} &    22:05:26.12  &  -05:44:31.1   &   0.86  &    0.0575 \\[-0.5ex]
{\rm A2457} &    22:35:40.81  &  +01:29:05.8   &   0.73  &    0.0584 \\[-0.5ex]
{\rm A2572a} &   23:17:11.95  &  +18:42:04.7   &   0.52  &    0.0390 \\[-0.5ex]
{\rm A2589} &    23:23:57.44  &  +16:46:38.3   &   0.95  &    0.0419 \\[-0.5ex]
{\rm A2593} &    23:24:20.08  &  +14:38:49.8   &   0.59  &    0.0417 \\[-0.5ex]
{\rm A2622} &    23:35:01.47  &  +27:22:20.9   &   0.55  &    0.0610 \\[-0.5ex]
{\rm A2626} &    23:36:30.49  &  +21:08:47.3   &   0.99  &    0.0548 \\[-0.5ex]
{\rm A2657} &    23:44:57.42  &  +09:11:35.2   &   0.82  &    0.0402 \\[-0.5ex]
{\rm A2665} &    23:50:50.55  &  +06:08:58.9   &   0.97  &    0.0556 \\[-0.5ex]
{\rm A2717} &    00:03:12.95  &  -35:56:13.3   &   0.52  &    0.0490 \\[-0.5ex]
{\rm A2734} &    00:11:21.64  &  -28:51:15.5   &   1.30  &    0.0625 \\[-0.5ex]
{\rm A3128} &    03:29:50.60  &  -52:34:46.8   &   2.71  &    0.0600 \\[-0.5ex]
{\rm A3158} &    03:43:29.69  &  -53:41:31.7   &   2.71  &    0.0593 \\[-0.5ex]
{\rm A3266} &    04:31:13.27  &  -61:27:11.9   &   3.14  &    0.0593 \\[-0.5ex]
{\rm A3376} &    06:00:41.09  &  -40:02:40.4   &   1.27  &    0.0461 \\[-0.5ex]
{\rm A3395} &    06:27:36.25  &  -54:26:57.9   &   1.43  &    0.0500 \\[-0.5ex]
{\rm A3490} &    11:45:20.15  &  -34:25:59.3   &   0.88  &    0.0688 \\[-0.5ex]
{\rm A3497} &    11:59:46.30  &  -31:31:41.6   &   0.74  &    0.0680 \\[-0.5ex]
{\rm A3528a} &   12:54:41.01  &  -29:13:39.5   &   0.68  &    0.0535 \\[-0.5ex]
{\rm A3528b} &   12:54:22.23  &  -29:00:46.8   &   1.01  &    0.0535 \\[-0.5ex]
{\rm A3530} &    12:55:35.99  &  -30:20:51.3   &   0.44  &    0.0537 \\[-0.5ex]
{\rm A3532} &    12:57:21.97  &  -30:21:49.1   &   1.44  &    0.0554 \\[-0.5ex]
{\rm A3556} &    13:24:06.71  &  -31:40:11.6   &   0.48  &    0.0479 \\[-0.5ex]
{\rm A3558} &    13:27:56.84  &  -31:29:43.9   &   3.20  &    0.0480 \\[-0.5ex]
{\rm A3560} &    13:32:25.76  &  -33:08:08.9   &   0.67  &    0.0489 \\[-0.5ex]
{\rm A3667} &    20:12:27.32  &  -56:49:36.3   &   4.47  &    0.0556 \\[-0.5ex]
{\rm A3716} &    20:51:56.94  &  -52:37:46.8   &   0.52  &    0.0462 \\[-0.5ex]
{\rm A3809} &    21:46:59.07  &  -43:53:56.2   &   1.15  &    0.0627 \\[-0.5ex]
{\rm A3880} &    22:27:54.43  &  -30:34:31.8   &   0.95  &    0.0584 \\[-0.5ex]
{\rm A4059} &    23:57:00.71  &  -34:45:32.8   &   1.58  &    0.0475 \\[-0.5ex]
{\rm IIZW108} &  21:13:55.90  &  +02:33:55.4   &   1.12  &    0.0483 \\[-0.5ex]
{\rm MKW3s} &    15:21:51.84  &  +07:42:32.1   &   1.37  &    0.0444 \\[-0.5ex]
{\rm RXJ0058} &  00:58:22.88  &  +26:51:52.6   &   0.22  &    0.0484 \\[-0.5ex]
{\rm RXJ1022} &  10:22:37.40  &  +38:34:45.0   &   0.18  &    0.0548 \\[-0.5ex]
{\rm RXJ1740} &  17:40:32.06  &  +35:38:46.1   &   0.26  &    0.0441 \\[-0.5ex]
{\rm ZwCl1261} & 07:16:41.24  &  +53:23:09.4   &   0.41  &    0.0644 \\ [-0.5ex]
{\rm ZwCl2844} & 10:02:36.54  &  +32:42:24.3   &   0.29  &    0.0503 \\[-0.5ex]
{\rm ZwCl8338} & 18:11:05.18  &  +49:54:33.7   &   0.40  &    0.0494 \\[-0.5ex]
{\rm ZwCl8852} & 23:10:42.27  &  +07:34:03.7   &   0.48  &    0.0408 \\[-0.5ex]
\hline
         \end{tabular}
      \]
        \tablecomments{The X-ray luminosity is shown in the 0.1-2.4 keV ROSAT RASS bandpass and it has been extracted from \cite{ebeling96}. The redshift information was taken from \cite{cava09}}
\label{tab:WINGS}
   \end{table}

 \begin{table}
      \caption[]{ACS BCGs sample}
      \[
        \begin{tabular}{lccccccl}
            \hline\noalign{\smallskip}
\multicolumn{1}{c}{\rm Name}&
\multicolumn{1}{c}{$\alpha$ (J2000)}&
\multicolumn{1}{c}{$\delta$ (J2000)}&
\multicolumn{1}{c}{$L_x$}&
\multicolumn{1}{c}{z}\\
\multicolumn{1}{c}{\rm }&
\multicolumn{1}{c}{hh:mm:ss}&
\multicolumn{1}{c}{dd:mm:ss}&
\multicolumn{1}{c}{10$^{44}$ erg/s}&
\multicolumn{1}{c}{}\\
\hline\noalign{\smallskip}           
{\rm RXJ0056.9-2740} &        00:56:56.1  &    -27:40:12 &    1.32 &   0.563 \\[0.01ex]
{\rm RXJ0110.3+1938} &        01:10:18.0  &     19:38:23  &   0.55  &  0.317 \\[0.01ex]
{\rm RXJ0154.2-5937} &        01:54:14.8  &    -59:37:48  &   1.25 &   0.360 \\[0.01ex]
{\rm RXJ0522.2-3625} &        05:22:14.2  &    -36:25:04  &   2.49 &   0.472 \\[0.01ex]
{\rm RXJ0826.1+2625} &        08:26:06.4  &     26:25:47 &    0.91 &   0.351 \\[0.01ex]
{\rm RXJ0841.1+6422} &        08:41:07.4  &     64:22:43  &   2.24 &   0.342 \\[0.01ex]
{\rm RXJ0847.1+3449} &        08:47:11.3  &     34:49:16 &    2.24 &   0.560 \\[0.01ex]
{\rm RXJ0926.6+1242} &        09:26:36.6  &     12:42:56 &    2.41 &   0.489 \\[0.01ex]
{\rm RXJ0957.8+6534} &        09:57:53.2   &    65:34:30 &    1.60 &   0.530 \\[0.01ex]
{\rm RXJ1015.1+4931} &       10:15:08.5  &     49:31:32  &   1.04  &  0.383 \\[0.01ex]
{\rm RXJ1117.2+1744} &       11:17:12.0  &     17:44:24 &    0.77 &   0.305 \\[0.01ex]
{\rm RXJ1123.1+1409} &       11:23:10.2  &     14:09:44 &    1.40 &   0.340 \\[0.01ex]
{\rm RXJ1354.2-0221} &       13:54:16.9  &     -02:21:47 &    2.52 &   0.546 \\[0.01ex]
{\rm RXJ1540.8+1445} &       15:40:53.3   &    14:45:34  &   0.96 &   0.441 \\[0.01ex]
{\rm RXJ1642.6+3935} &       16:42:38.9  &     39:35:53  &   0.86 &   0.355 \\[0.01ex]
{\rm RXJ2059.9-4245} &      20:59:55.2  &    -42:45:33 &    0.81&    0.323 \\[0.01ex]
{\rm RXJ2108.8-0516} &       21:08:51.2  &     -05:16:49  &   0.81 &   0.319 \\[0.01ex]
{\rm RXJ2139.9-4305} &       21:39:58.5  &    -43:05:14  &   0.79 &   0.376 \\[0.01ex]
{\rm RXJ2202.7-1902} &       22:02:44.9  &    -19:02:10 &    0.82 &   0.438 \\[0.01ex]
{\rm RXJ2328.8+1453} &       23:28:49.9  &     14:53:12  &   1.16 &   0.497 \\[0.01ex]
\hline
         \end{tabular}
      \]
      \tablecomments{The X-ray luminosity is in the 0.5Ð2.0 keV energy band. Both X-ray luminosity and redshifts have been extracted from \cite{mullis03} and references herein.}
\label{tab:ACS}
   \end{table}

\section[]{Surface brightness analysis}

We analyzed the surface brightness distribution of the galaxies by using GASP2D \citep{mendezabreu08,ascaso09}. This routine fits the 2D surface brightness distribution of galaxies with one or two components following a particular surface brightness model. In particular, we have fit the surface brightness of the galaxies with two components: S\'ersic  \citep{sersic68} and  exponential \citep{sersic68,freeman70}.

All the information regarding GASP2D can be found in \cite{mendezabreu08}. Here, we will only mention some important remarks. GASP2D fits individually each galaxy. It first masks the rest of the galaxies in the frame automatically. After that, the user is allowed to modify them in case some galaxies have not been correctly deblended or detected.

GASP2D adopts a Levenberg-Marquardt algorithm to fit the two-dimensional surface brightness distribution of the galaxy. Since the fitting algorithm is based on the $\chi^2$ minimization, it is important to start the procedure adopting initial trials for the free parameters as close as possible to their actual values. These initial conditions were obtained by fitting the one-dimensional surface brightness  ellipticity and position angle isophotal profiles of the galaxy.  The routine works out the best initial conditions by fitting an exponential law at large radii and a bulge (usually S\'ersic or de Vaucouleur) model to the residual surface brightness profile that results at subtracting the outer fit component to the overall profile. This ensures that the iteration procedure does not just stop on a local minimum of the $\chi^2$ distribution. In addition, during each iteration of the fitting algorithm the seeing effects were taken into account by convolving the model image with a  Moffat point spread function (PSF) with the fast Fourier transform algorithm. The PSF FWHM matches the one measured from the foreground stars in the field. The code also allows to introduce a Gaussian or a star image to reproduce the PSF.

It has been a wide discussion in the literature about the optimum number of components to fit the surface brightness of a BCG. A number of works \citep{caon93,graham96,patel06} argued that a much better model to fit the surface brightness of the BCG comes from a S\'ersic profile rather than a de Vaucouleurs profile, since the universality of the latter is uncertain. However, many recent works have shown evidence that the BCG outermost regions can not be described by  a S\'ersic model and to provide a satisfactory fit, it is necessary  the introduction of at least two components \citep{nelson02,gonzalez03,gonzalez05,seigar07,liu08}.

Motivated by the fact that at least two components are necessary to fit the main BCG population, we have decided to fit the whole population of BCGs with two S\'ersic+Exponential components \citep{donofrio01,seigar07,vikram09}. Then, we will call effective radius the effective radius obtained from the S\'ersic component.  Even if this value does not correspond exactly with the effective radius of the whole galaxy, we have studied the model dependence of the results  (see section 6.1) and we have found the results to be robust.

\begin{figure*}
\centering
\includegraphics[clip,angle=0,width=.8\hsize]{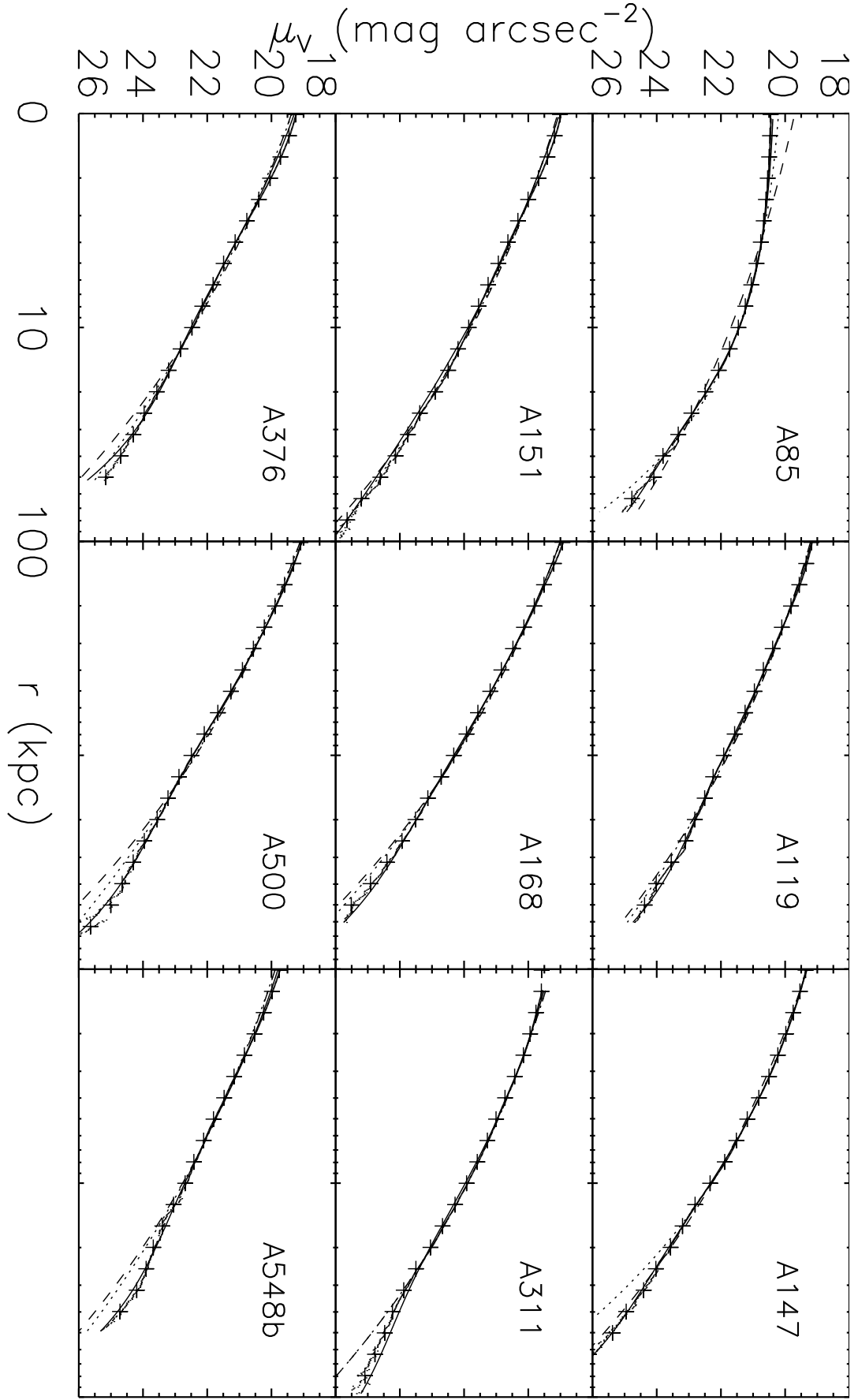}
\caption{One dimensional surface brightness profiles for the first nine BCGs in the WINGS sample (crosses). We overplot the  two component S\'ersic+Disc fit (solid line), the single S\'ersic fit (dashed line) and the single De Vaucouleur fit (dotted line). The axis scale is the same for all the plots.}
\label{fig:componentsWINGS}
\end{figure*}

\begin{figure*}
\centering
\includegraphics[clip,angle=0,width=.8\hsize]{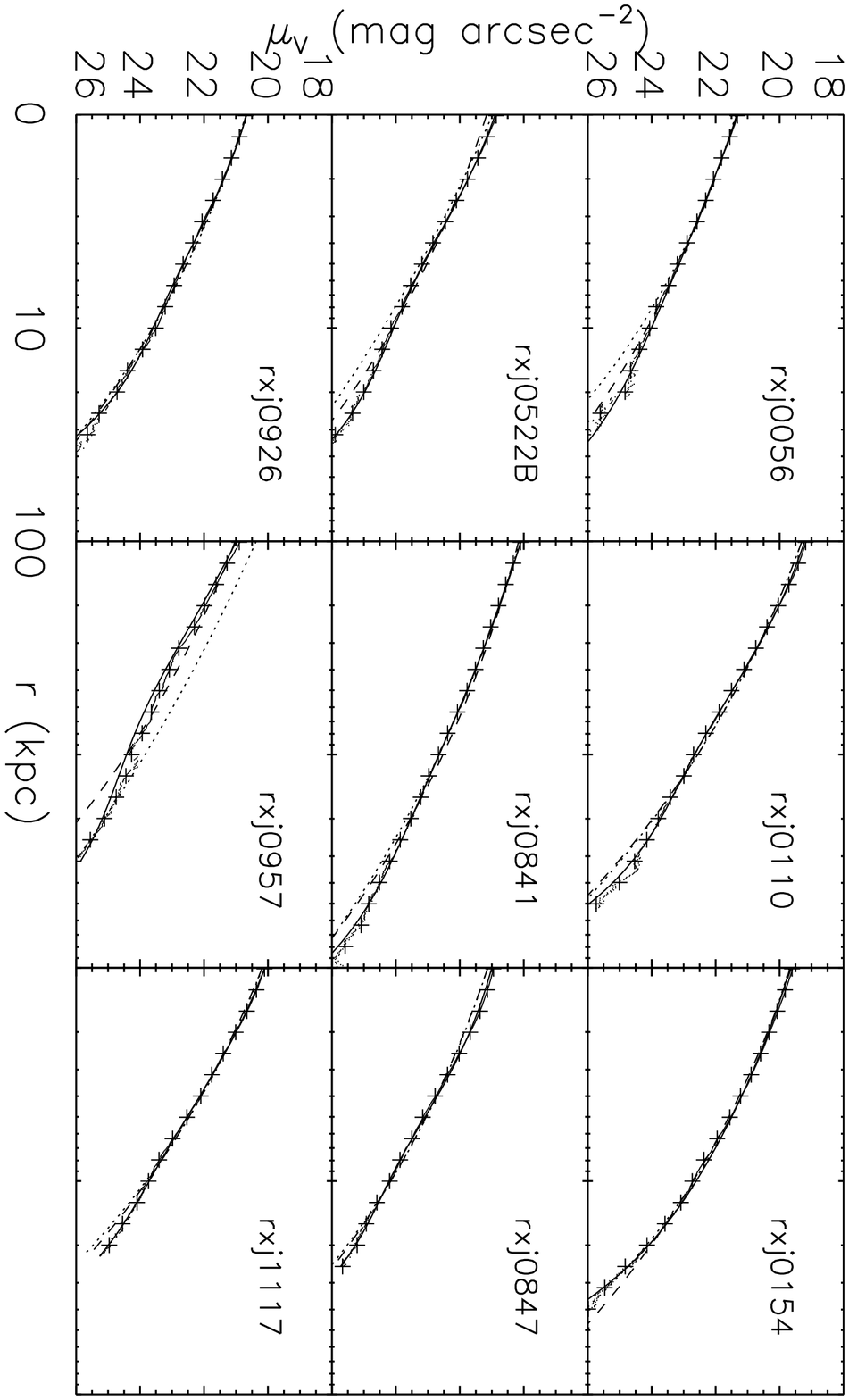}
\caption{One dimensional surface brightness profiles for the first nine BCGs in the ACS sample (crosses). We overplot the  two component S\'ersic+Disc fit (solid line), the single S\'ersic fit (dashed line) and the single De Vaucouleur fit (dotted line). The axis scale is the same for all the plots.}
\label{fig:componentsACS}
\end{figure*}

In Figures \ref{fig:componentsWINGS} and \ref{fig:componentsACS}, we show some examples of the one-dimensional surface brightness of the BCGs in both samples together with the overlapped fits of the S\'ersic and Exponential model (solid line), S\'ersic model (dashed line) and De Vaucouleur model (dotted line). All the fits have been performed up to  25 mag arcsec$^{-2}$ in the V band. We list the results for each sample of the S\'ersic+Exponential, S\'ersic and De Vaucouleur fits in Appendix A, B and C respectively. Note that the S\'ersic+Exponential fits get the best $\chi^2$ values compared to the single component fits.

The sample of five BCGs analyzed in \cite{seigar07} is much deeper than our two BCG samples. They have a surface brightness limit of  27.5-28 for their sample, while we arrive down to $\mu_V$=25.  Our objective in this paper is to compare the 'sizes' and 'concentration' of two different BCGs samples. We have ensured that the results in both samples are consistent, since the resolution for both samples is similar and we arrive up to the same surface brightness limit and use the same procedure.

\section{Structural parameters}

In this section, we only have considered the BCGs from the WINGS sample with  $5 \times 10^{43}<L_x<2.52 \times 10^{44} \,erg \,s^{-1}$ in order to match the same X-ray luminosity range as the ACS cluster sample. We have analyzed the structural parameters extracted from the surface brightness analysis for the BCGs for the WINGS and the ACS samples. 

\subsection{Sizes and shapes}

In Figure \ref{fig:nre}, we show the relation between the S\'ersic parameter ($n$) and the effective radius ($r_e$) for the BCGs from the WINGS (black points) and the ACS (diamonds) samples. In both cases, we see a trend in the sense that larger BCGs have larger S\'ersic parameter. This relation has
also been observed for bright elliptical galaxies in nearby galaxy clusters \citep{caon93,graham03,aguerri04}. The triangle and the square in Figure \ref{fig:nre} show the median values of $\log(n)$ and $\log(r_{e})$ for the BCGs from the WINGS and the ACS samples, respectively. The linear fits of these relations are given by: 

\begin{figure}
\includegraphics[clip,width=1.\hsize]{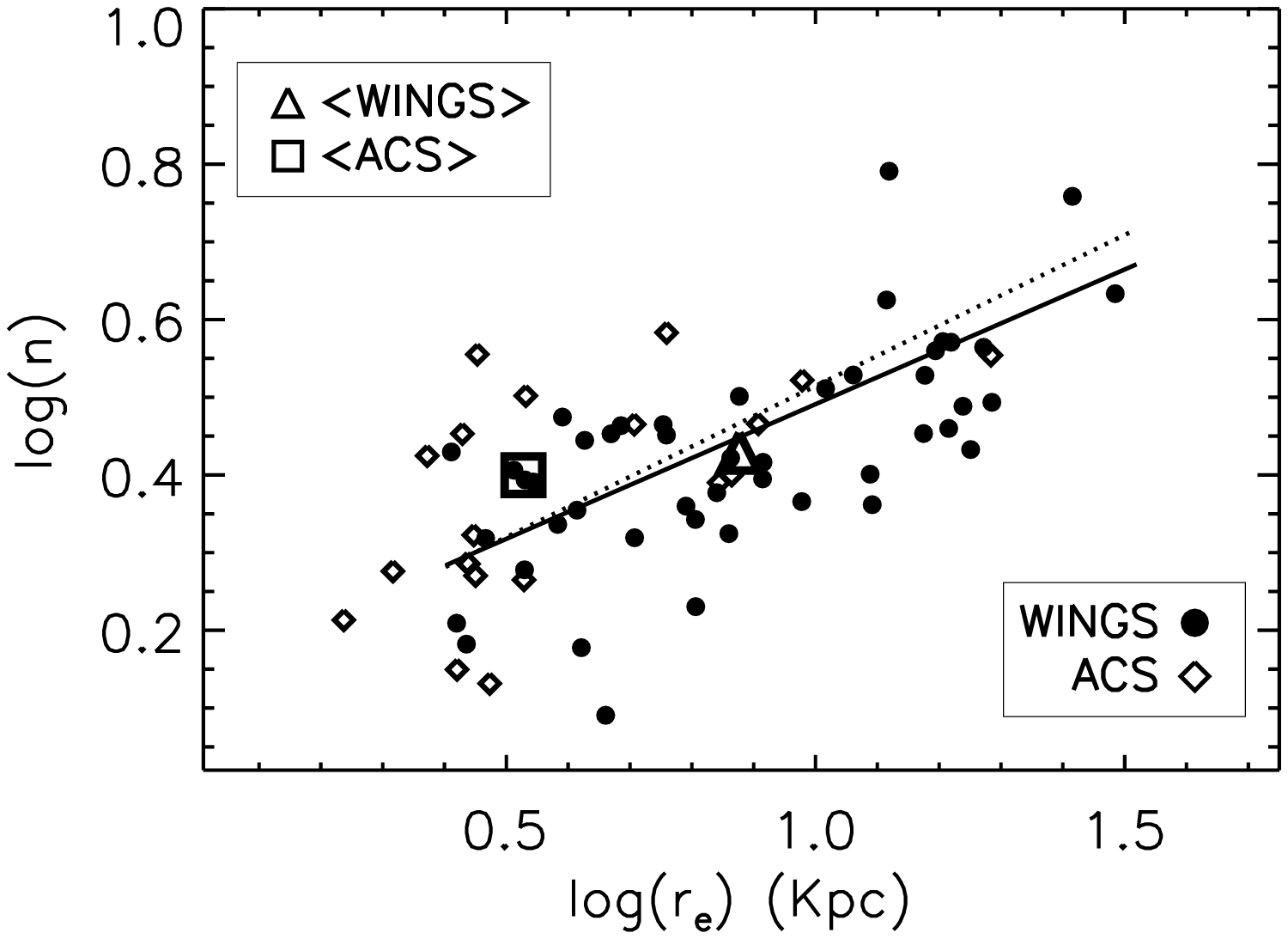}
\caption{Relationship between $\log(r_e)-\log(n)$ for the BCGs in WINGS (black points) and ACS (diamonds). The solid and dotted lines show the fits to the WINGS and ACS samples respectively. The triangle and square show the median value for the WINGS and ACS sample respectively.}
\label{fig:nre}
\end{figure}

\begin{equation}
\log n = (0.144\pm 0.018)  + (0.347 \pm 0.018) \log r_e
\end{equation}

\begin{equation}
\log n = (0.126 \pm 0.018)  + (0.389 \pm 0.031) \log r_e
\end{equation}
for the WINGS and the ACS samples, respectively. These fits have been obtained by using a 3$\sigma$ clipping algorithm. They are also overplotted in Figure \ref{fig:nre} with solid (WINGS) and dotted (ACS) lines. Notice that the slopes are similar within the errors.
  
In Table \ref{tab:params}, we show the median values for the shape parameters, effective radius and mean surface brightness for both samples. The errors  have been estimated with a bootstrap algorithm. While both samples have very similar values of the S\'ersic parameter ($n(z=0)/n(z\sim0.5)=1.05\pm0.14$), we do see a difference for the effective radius between both samples. Thus, nearby BCGs are  larger than intermediate redshift ones, being $r_e(z=0)/r_e(z\sim0.5)$= 2.06$\pm$0.63. We have performed a KS test resulting that the galaxy sizes distributions of the nearby and intermediate redshift samples are statistically different. In contrast, the S\'ersic parameter distributions of both galaxy samples are not statistically different. The fact that the S\'ersic parameter of the BCGs has not  changed indicates that the central light concentrations of the BCGs are similar in both samples.

\begin{table}
      \caption{Shapes and Sizes for BCGs Samples}
      \[
         \begin{array}{lll}
            \hline\noalign{\smallskip}
\multicolumn{1}{c}{\rm }&
\multicolumn{1}{c}{\rm WINGS (z\sim0) }&
\multicolumn{1}{c}{\rm ACS (z\sim0.5)}\\
$ $ &$ $ &$ $  \\
\hline\noalign{\smallskip}
{\rm <n>} & 2.64 \pm 0.12 &  2.51 \pm 0.32  \\
{\rm <r_e (kpc)>} & 6.92 \pm 1.40 & 3.35 \pm 0.77 \\
{\rm <\mu_e (mag/arcsec^{2})>} & 20.29 \pm 0.23  & 20.96 \pm 0.26  \\
$ $ &$ $ &$ $  \\
\hline
         \end{array}
      \]
\label{tab:params}
   \end{table}

\subsection{Kormendy relation}

We have also fitted the Kormendy relation \citep{kormendy77} for both samples as shown in Figure \ref{fig:kormendy}.  In this relation,  \mbox{$<\mu_e>$} refers to the median effective surface brightness within $r_e$. The linear fits are given by:
  
\begin{figure}
\includegraphics[clip,width=1.\hsize]{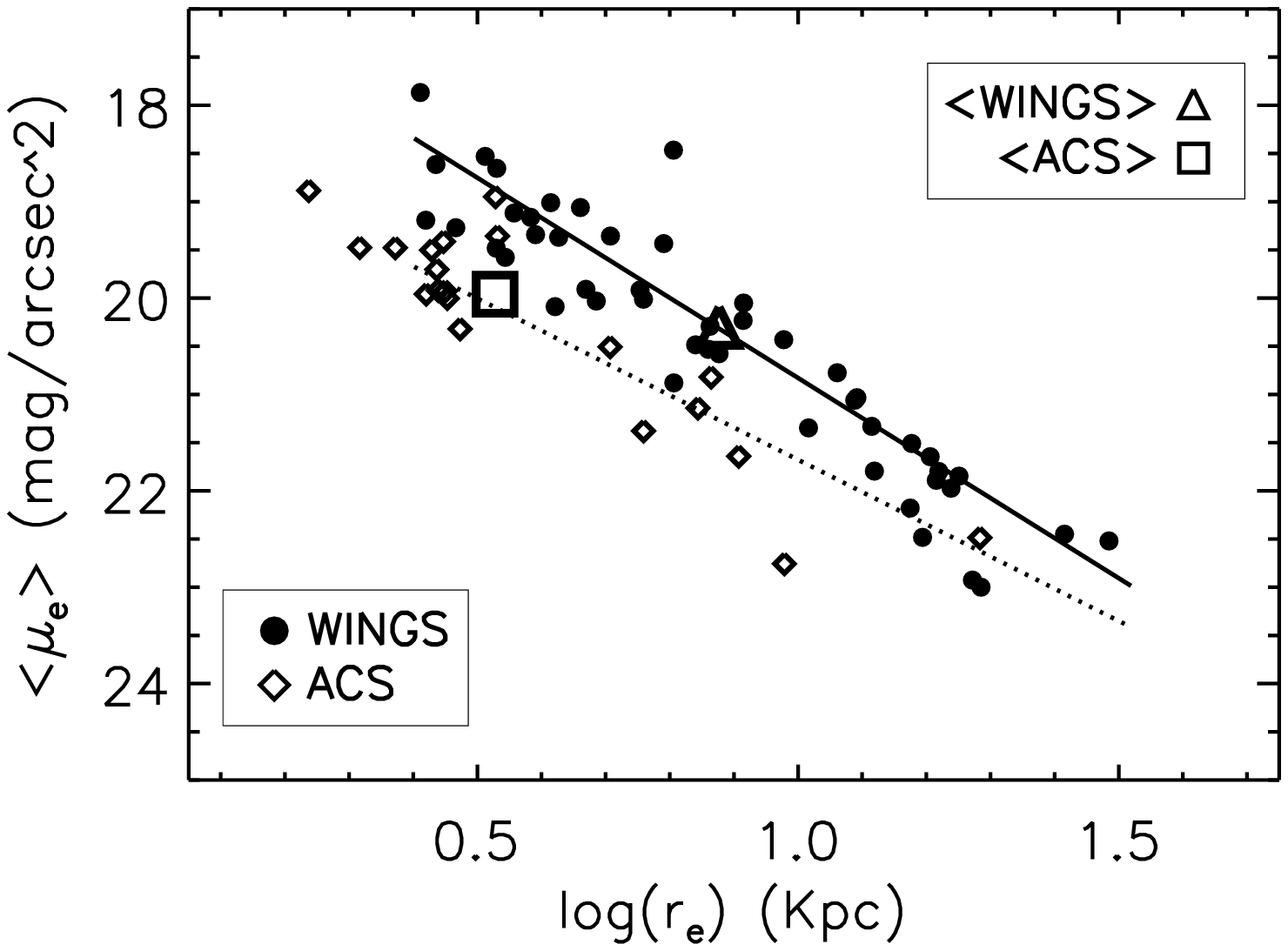}
\caption{Kormendy relation for the BCGs in WINGS (black points) and ACS (diamonds). The solid and dotted lines show the fits to the WINGS and ACS samples respectively. The triangle and square show the median value for the WINGS and ACS samples respectively.}
\label{fig:kormendy}
\end{figure}

\begin{equation}
<\mu_e > = (16.675\pm 0.184)  + (4.154 \pm 0.209) \log r_e
\end{equation}

\begin{equation}
<\mu_e > = (18.332 \pm 0.161)  + (3.346 \pm 0.253) \log r_e 
\end{equation}
     
for the WINGS and the ACS samples, respectively. These fits have also been performed with a 3$\sigma$ clipping algorithm. We have obtained a different Kormendy relation for the different samples with a much steeper slope for the local sample. The median values (see Table \ref{tab:params}) show that the intermediate redshift BCGs are smaller and have similar effective surface brightness than  low redshift ones. Indeed, the KS test show that the mean surface brightness distributions are not statistically different for the WINGS and ACS samples. Our relations agree with the  \citep{bildfell08}. They recently found a steeper slope ($\sim$ 3.96) for the Kormendy relation for BCGs at 0.15 $\le$ z $\le$ 0.55 compared with local ellipticals.

\subsection{Structural parameters versus luminosity}

\begin{figure}
\centering
\includegraphics[clip,width=1.\hsize]{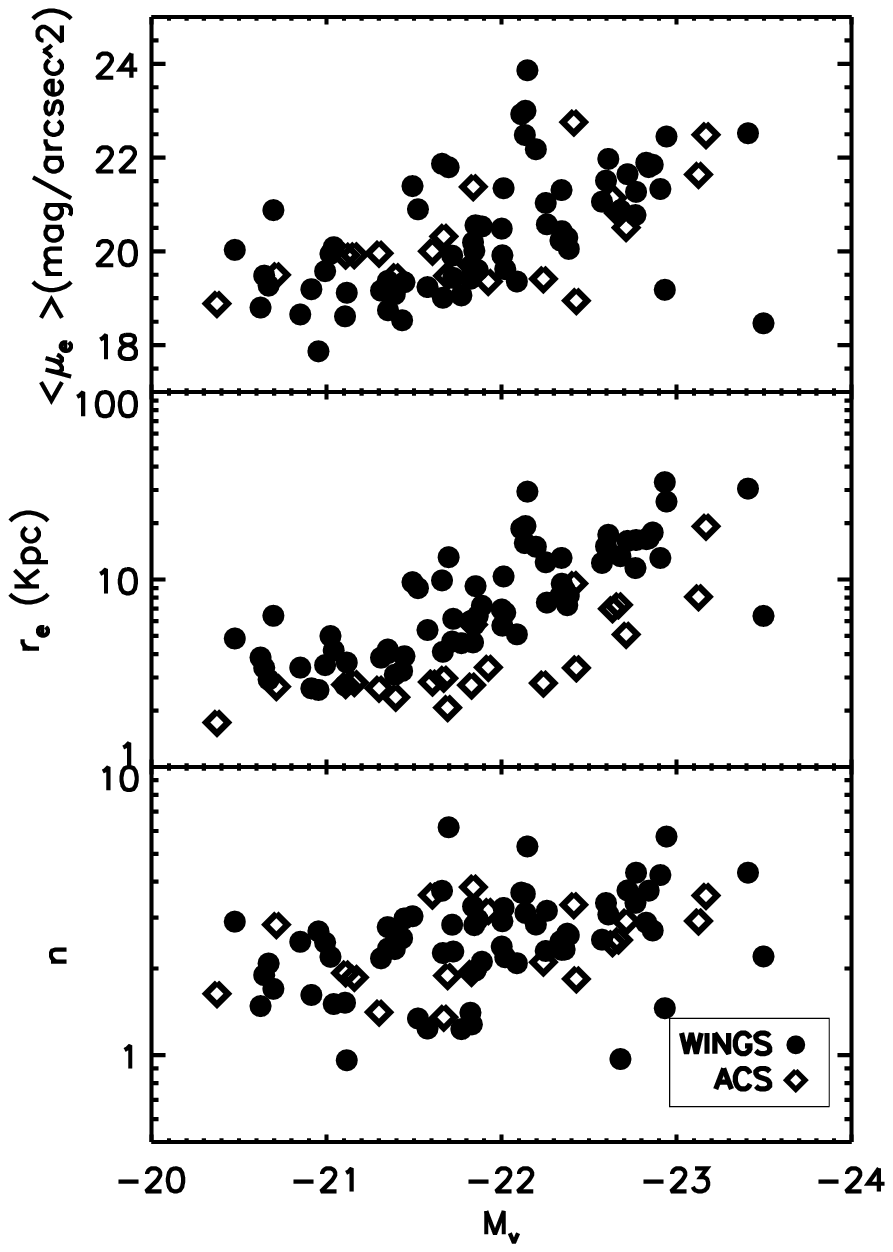}
\caption{Absolute magnitude versus medium surface brightness, S\'ersic parameter and effective radius for the BCGs in WINGS (black points) and ACS (diamonds) for the S\'ersic component from the two components model fit.}
\label{fig:mags}
\end{figure}

In Figure  \ref{fig:mags}, we show the relation between absolute V rest-frame magnitude (M$_{V}$) of the fitted S\'ersic component and the mean surface brightness, S\'ersic parameter and effective radius of the BCGs. For a given luminosity, nearby BCGs have fainter $< \mu_{e} >$, larger $r_e$ and similar S\'ersic parameter than the intermediate redshift BCGs sample as it is shown in Table \ref{tab:params}.

Notice also the same behavior in the two BCGs samples. Thus, brighter BCGs are larger (the Spearman test provides a significance level of 6.81 and 3.56 $\sigma$ for the WINGS and the ACS sample respectively), having the lower redshift sample a steeper slope with respect to the intermediate redshift sample. The linear fits are given by:

\begin{equation}
\log r_e = (-6.966 \pm 0.323)  + (-0.356\pm 0.015) M_V
\end{equation}

\begin{equation}
\log r_e = (-5.057 \pm 0.316)  + (-0.260 \pm 0.015) M_V
\end{equation}

for the WINGS and ACS sample respectively. The slopes of the local sample agrees with other works  \citep{bernardi07}. %vonderlinden07,

This size-luminosity relation have also been supported for early-type galaxies by \cite{caon93,gutierrez04,aguerri05,liu08,bernardi09}. BCGs in  low redshift clusters  have also a significant correlation between absolute magnitude and shape parameter (3.65$\sigma$ significance in the Spearman test) and between absolute magnitude and mean effective surface brightness (4.79$\sigma$ significance in the Spearman test).  However, these tendencies are less significant for the intermediate redshift sample.

\section{Relationship between BCGs and their host cluster.}

We have investigated any relationship between the global parameters of the host cluster and the structural parameters of the BCGs. Since most of the information about the host cluster for the ACS sample is not available, we have only considered the WINGS sample. These relations will help to constrain theories of formation and evolution of the clusters and the BCGs themselves.  We have considered three different global parameters for the clusters: X-ray cluster luminosity ($L_x$), the degree of dominance ($\Delta m$; \citealt{kim02}), and the distance between the X-ray peak and the BCG center (D). These parameters indicate different global properties of galaxy clusters. 

It is well known that the X-ray cluster luminosity correlates with the temperature of the hot gas present in galaxy clusters \citep{vikhlinin05} and that there is a physical relation between hot gas temperature and mass of the galaxy cluster  \citep{finoguenov01,vikhlinin06} . Therefore, $L_{X}$ is an indication of the mass of the cluster \citep{reiprich02}.

The degree of dominance is defined as the difference between the magnitude of the BCG and the mean magnitude of the second and third brightest galaxies of the cluster within the central 500 kpc. It is an indicator of how dominant the BCG is with respect to the cluster. In the hierarchical scenario, the natural evolution of galaxy clusters is to accrete mass to the center of the cluster were the BCGs are located. In other words, clusters with larger $\Delta m$ would be more evolved systems. The extreme cases are the galaxy fossil clusters or groups \citep{ponman94}.

On the other hand, a good indicator of the dynamical state of the galaxy cluster is the closeness of the X-ray center of the cluster and the position of the BCG   \citep{collins03,shan10}.

\begin{figure*}
\centering
\includegraphics[clip,angle=0,width=1.\hsize]{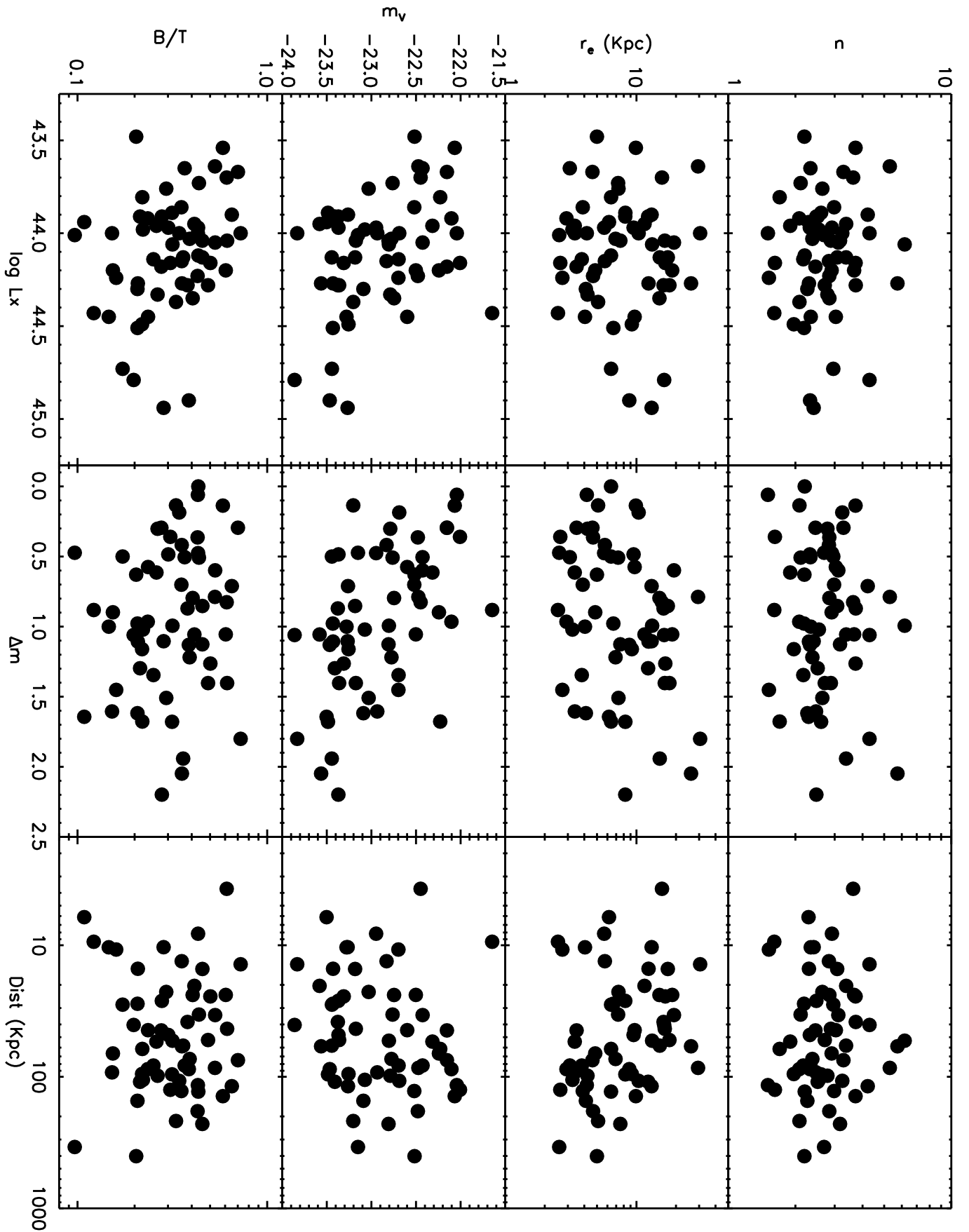}
\caption{Structural parameters of the BCGs (shape parameter, effective radius, ellipticity, absolute magnitude and bulge-to-total light fraction) versus different properties of the host clusters ($L_x$,  degree of dominance and distance from the X-ray center) for the BCGs in WINGS sample.}
\label{fig:host}
\end{figure*}

In Table \ref{tab:corr}, we list the significance of the Spearman correlation test for the different relations and in Figure \ref{fig:host}, we show the different relationships for the WINGS BCG sample. We find significant correlations between the cluster X-ray luminosity and the BCGs absolute magnitude and the B/T parameter. Thus, more luminous X-ray clusters host BCG with smaller values of B/T, showing that as the cluster becomes more massive, the luminosity of the internal regions of the BCG contributes less to their total light. One possible interpretation is that brighter envelopes are located in BCGs placed in the most X-ray luminous clusters. This behavior is in agreement with previous works \citep{vandokkum10,liu09}. In addition, there is a significant tendency of finding more luminous BCGs in more X-ray luminous clusters, consistent with optical-X-ray luminous function  \citep{lin04,popesso06}. If we assume that light trace mass  \citep{reyes08}, this would imply that BCGs are also more massive in more X-ray luminous clusters as observed in other works, \citep{burke00,stott08}.

\begin{table}
      \caption{Significance of the Spearman Test for the the structural parameters of the BCGs and their host cluster properties in WINGS sample.}
      \[
         \begin{array}{lccc}
            \hline\noalign{\smallskip}
\multicolumn{1}{c}{\rm }&
\multicolumn{1}{c}{\rm log \, L_x }&
\multicolumn{1}{c}{\rm log \, \Delta m }&
\multicolumn{1}{c}{\rm log \, D}\\
\hline\noalign{\smallskip}
\rm log \, n       &    0.66   &   -0.27  &     0.89 \\
\rm log \, r_e       & -0.49 &    -2.26 &     2.32  \\
\rm mag  &                2.22 &      3.48 &    -1.53  \\
\rm log \, B/T       &    2.27 &   1.41 &     0.06\\
\hline\noalign{\smallskip}
         \end{array}
      \]
\label{tab:corr}
   \end{table}

On the other hand, we do not find a significant correlation between $\Delta m$ and shape parameter. In contrast, significant correlations are found between $\Delta m$ and magnitude and effective radius. Thus, BCGs located in clusters with larger  degree of dominance are larger and more luminous. These results are in agreement with \citealt{bildfell08,niederste-ostholt10} and \citealt{smith10}. The larger and brighter BCGs in massive clusters suggest the evolutionary processes implied in transforming the BCGs are also transforming their host clusters.

The significance of the correlations between the structural parameters and the distance to the center of the cluster are shown in the last column in Table  \ref{tab:corr}. We do see a significant correlation between the location of the BCG in the cluster and the effective radius of the BCG. Thus, larger galaxies tend to be closer to the center of the cluster potential well given by X-ray data displaying a more dynamically evolved stage in the cluster. This result agrees with the results obtained from a X-ray analysis of an intermediate redshift cluster sample by \cite{sanderson09}.

As a conclusion, the properties of the BCGs and their host galaxy clusters are closely related. In particular, larger and brighter BCGs, with smaller B/T are located near to the center of the potential well of very luminous and dominant clusters. This points to a connection between the BCG formation processes and the mass assembly in galaxy clusters. Dynamically evolved and massive galaxy clusters are hosting more massive BCGs, with larger halos, suggesting that the processes happening in the cluster are more active in denser environments.

 \section{Discussion and Conclusions}

In the present work, we have performed an analysis of two BCG samples at different redshift ranges. On one hand, we have analyzed the evolution of their structural parameters and on the other, the relation between BCGs and their host galaxy cluster for the WINGS sample. We discuss here the robustness of the results and their implications on the formation and evolution of these particular galaxies.

\subsection{Robustness of the results}

We have fitted the surface brightness distribution of BCGs with a two components model: S\'ersic plus exponential. We have observed that the median values of the structural parameters of the S\'ersic fitted component have evolved during the last 6 Gyr. In particular, the effective radius has changed by $r_{e}(z\sim0)/r_{e}(z\sim0.5)=2.06\pm0.63$. In contrast, the shape S\'ersic parameter does not change, being $n(z\sim0)/n(z\sim0.5)=1.05\pm0.14$. But, how does these results depend on the fitted model? In order to answer to this question, we have also fitted the surface brightness distribution of our galaxies with a single S\'ersic and de Vaucouleurs profiles.

Independent of the fitted model, there is not variation within the last 6 Gyrs in the S\'ersic shape parameter. In the case of a single S\'ersic fit we have obtained $n(z\sim0)/n(z\sim0.5)$=1.02$\pm$ 0.21. In contrast, the rate  of variation in the size depends on the fitted model. Thus, the change of $r_{e}$, when only one single S\'ersic component was fitted, becomes by $r_{e}(z\sim0)/r_{e}(z\sim0.5)=$ 1.89$\pm$ 0.36 and for a single de Vaucouleur fit, we find  $r_{e}(z\sim0)/r_{e}(z\sim0.5)=$ 1.47$\pm$ 0.23. This implies that the growth size rate of the galaxies is smaller when only one component was fitted. However, one single component fits (S\'ersic and de Vaucouleurs) give much worse $\chi^2$ values than a S\'ersic+ Exponential.

From a model independent perspective, we have measured the size of the galaxies in a different way. We have calculated the 'global' effective radius of BCGs by solving the equation: $L(<r_e) =L_{total}/2$, being $L_{total}$ the total integrated luminosity of the galaxy. In this case, we have obtained that  $r_{e}(z\sim0)/r_{e}(z\sim0.5)=1.70\pm0.15$. 

Thus, the effective radius growth extracted from a model independent measurement is smaller but consistent with the growth obtained by using a two component fitting model or a single S\'ersic model. Thus, BCGs at z$\sim$0.5 are  smaller than nearby ones with independence of the procedure we use to calculate the sizes.

\subsection{Evolution of BCGs during the last 6 Gyrs}

There are several observational pieces of evidence about the fact that massive early-type galaxies have grown in size from z$\sim 2$ \citep{daddi05,trujillo06,trujillo07,vandokkum10}.  However, \cite{mancini10} have presented discrepant results by showing that some high redshift massive elliptical have similar sizes to local ones. As long as BCGs are concerned, \cite{nelson02}  and   \cite{bernardi09} reported an increase in the sizes of BCGs at intermediate redshift compared with local ones.  \cite{nelson02} informed on a factor of $\sim$1.7 since z $\sim$0.25 while  \cite{bernardi09} published a factor of $\sim$2 since z $\sim$0.5.

Recently, it has been discovered than the S\'ersic shape parameters of early-type galaxies has also evolved during the last Gyrs, being larger for nearby galaxies \citep{vikram09,vandokkum10}. Indeed, although the mass of massive early-type galaxies has grown from z$\sim2$ until today, this mass growth has been focused on their external regions \citep{vandokkum10}. These results have been interpreted as an inside-out growth of the early-type galaxies, assembling their extended haloes in the last Gyrs. There are several numerical simulations supporting those observed changes of the structural parameters of early-type galaxies. Thus, major or minor mergers produce a growth of the effective radius and S\'ersic parameter of the galaxy \citep{aguerri01,scannapieco03,eliche-moral06,hopkins10}.

The results presented in this work show that BCGs have grown in size within the last 6 Gyrs by a factor of $\sim$ 2. In addition, the growth rate is similar making use of a model independent measurement  such as the 'global' effective radius calculated from the whole luminosity. The difference between the evolution of the BCGs and other massive early-type galaxies is the constancy of the S\'ersic shape parameter in BCGs since z $\sim$0.6.  The fact that intermediate redshift and nearby BCGs show no evolution in the S\'ersic parameter implies that the evolution of these galaxies in the last 6 Gyrs has not been driven by galaxy mergers because major or minor mergers would have changed the shape of the surface brightness distribution of the galaxies.

Numerical simulations show that the structural parameters of early-type galaxies can change due to several processes. In particular, if a galaxy could lose a fraction of its central mass then the radius of the object will grow, and the system will keep the surface brightness profile shape \citep{hopkins10}. This process called adiabatic expansion could explain our observables for BCGs. The loss of the inner material could be due to different reasons. Among others, quasar feedback can produce a loss of a considerable fraction of baryonic matter in the center of galaxies \citep{fan08}. Central starburst, produced by cooling flows observed in some BCGs \citep{fabian82}  can also activate galactic winds and superwinds and eject part of the inner mass in galaxies \citep{tenorio-tagle05,silich10}. Displacement of black holes from galaxy center transfer energy to stars in the nucleus and can convert density cusp profiles in core ones. This would also produce an enlarge of the system \citep{merritt04}.

In summary, according to our observations, we conclude that the only mechanisms that are able to explain the BCGs evolution in size but not in shape during the last 6 Gyrs are feedback processes. Thus, the evolution of BCGs within the last 6 Gyrs is driven by feedback processes rather than merger evolution. This result is in contradiction with the results obtained by recent numerical simulations about the origin and evolution of BCGs  \citep{deLucia07b}. These simulations predict an important mass growth of the galaxies via dry mergers in the last 6 Gyrs. Nevertheless, other observational recent works have also observed a small or negligible change of the mass of BCGs in the last 8 Gyrs \citep{collins09,stott10}.

\acknowledgments

We thank the anonymous referee for the valuable comments that improved this paper. We acknowledge Andrea Biviano, Bianca Poggianti and Gianni Fasano for helpful comments. We also thank Alfonso Arag\'on-Salamanca and Anthony Gonzalez for stimulating and helpful discussion. Special thanks to Dave Wittman. JALA has been founded by the Spanish MICINN under the AYA2010-21887-C04-04 project.  BA acknowledges the support of NASA grant NNG05GD32G.

{\it Facilities:} \facility{INT}, \facility{HST (ACS)}, \facility{MPG-ESO}

\clearpage

\appendix

\section{Results of the 2-components S\'ersic+Exponential fit}

\subsection[]{WINGS BCGs sample}

\begin{longtable}{lcccccccccccc}
            \hline
\multicolumn{1}{c}{\rm Name}&
\multicolumn{1}{c}{$\mu_e$}&
\multicolumn{1}{c}{$r_e$}&
\multicolumn{1}{c}{$e_b$}&
\multicolumn{1}{c}{$\mu_0$}&
\multicolumn{1}{c}{$h$}&
\multicolumn{1}{c}{$e_d$}&
\multicolumn{1}{c}{$n$}&
\multicolumn{1}{c}{$PA_b$}&
\multicolumn{1}{c}{$PA_d$}&
\multicolumn{1}{c}{$B/T$}&
\multicolumn{1}{c}{$m_V$}&
\multicolumn{1}{c}{$\chi^2$}\\
\multicolumn{1}{c}{}&
\multicolumn{1}{c}{mag/arc sec$^2$}&
\multicolumn{1}{c}{kpc}&
\multicolumn{1}{c}{}&
\multicolumn{1}{c}{mag/arc sec$^2$}&
\multicolumn{1}{c}{kpc}&
\multicolumn{1}{c}{}&
\multicolumn{1}{c}{}&
\multicolumn{1}{c}{}&
\multicolumn{1}{c}{}&
\multicolumn{1}{c}{}&
\multicolumn{1}{c}{}&
\multicolumn{1}{c}{}\\
\hline\noalign{\smallskip}
A85 &        17.99 &        13.06 &         0.80 &        18.56 &        35.54 &         0.57 &         0.97 &          53 &          60 &         0.30 &        13.21 &         2.31 \\ [0.01ex]
A119 &        17.63 &         6.59 &         0.88 &        18.02 &        25.34 &         0.61 &         2.19 &         123 &         128 &         0.21 &        13.04 &         3.24 \\ [0.01ex]
A147 &        17.79 &         7.25 &         0.76 &        19.27 &        26.56 &         0.73 &         2.11 &         150 &         137 &         0.44 &        13.72 &         2.20 \\ [0.01ex]
A151 &        19.79 &        29.94 &         0.80 &        21.20 &        68.51 &         1.00 &         4.30 &         163 &          12 &         0.72 &        13.04 &         0.43 \\ [0.01ex]
A168 &        17.95 &         7.54 &         0.94 &        19.02 &        24.28 &         0.51 &         3.17 &          63 &          59 &         0.46 &        13.69 &         2.12 \\ [0.01ex]
A311 &        18.43 &        12.23 &         0.77 &        19.47 &        64.55 &         0.24 &         2.52 &         110 &         122 &         0.21 &        13.96 &        28.39 \\ [0.01ex]
A376 &        17.21 &         3.79 &         0.93 &        18.06 &        15.93 &         0.81 &         2.17 &           6 &           4 &         0.25 &        13.92 &         2.94 \\ [0.01ex]
A500 &        18.29 &         5.73 &         0.78 &        19.55 &        24.19 &         0.63 &         2.83 &           6 &          36 &         0.35 &        14.59 &         0.45 \\ [0.01ex]
A548b &        18.70 &         4.87 &         0.91 &        19.17 &        19.64 &         0.69 &         2.19 &          16 &         131 &         0.20 &        13.80 &         0.67 \\ [0.01ex]
A602 &        20.13 &        18.93 &         0.94 &        20.19 &        32.77 &         0.32 &         3.12 &          62 &          72 &         0.53 &        14.79 &        10.06 \\ [0.01ex]
A671 &        18.21 &        11.39 &         0.80 &        18.65 &        30.29 &         0.63 &         3.38 &         115 &         120 &         0.41 &        13.19 &         5.45 \\ [0.01ex]
A754 &        17.70 &         8.81 &         0.67 &        18.29 &        24.30 &         0.75 &         2.33 &          18 &          29 &         0.39 &        13.47 &         4.21 \\ [0.01ex]
A957 &        17.66 &         8.05 &         0.79 &        18.31 &        27.49 &         0.64 &         2.61 &          11 &         152 &         0.32 &        13.01 &        10.21 \\ [0.01ex]
A970 &        17.43 &         3.46 &         0.93 &        18.51 &        15.51 &         0.68 &         2.46 &          23 &          43 &         0.28 &        14.96 &         2.95 \\ [0.01ex]
A1069 &        16.62 &         3.19 &         0.74 &        17.49 &        15.27 &         0.74 &         2.55 &         180 &           0 &         0.22 &        14.27 &        13.04 \\ [0.01ex]
A1291 &        21.13 &        29.13 &         0.81 &        19.83 &        30.62 &         0.32 &         5.30 &         168 &         171 &         0.53 &        14.30 &         5.45 \\ [0.01ex]
A1631a &        17.72 &         3.84 &         0.76 &        18.39 &        12.50 &         0.70 &         2.98 &         145 &         148 &         0.35 &        14.03 &         0.67 \\ [0.01ex]
A1644 &        18.28 &         3.82 &         0.83 &        18.01 &        17.77 &         0.60 &         1.48 &          51 &          42 &         0.08 &        13.47 &         3.61 \\ [0.01ex]
A1668 &        20.09 &        18.36 &         0.87 &        18.78 &        15.12 &         0.57 &         3.67 &         153 &         167 &         0.60 &        14.77 &         0.90 \\ [0.01ex]
A1736 &        17.36 &         5.20 &         0.62 &        18.49 &        20.34 &         0.55 &         2.09 &         137 &         132 &         0.33 &        13.34 &         0.71 \\ [0.01ex]
A1795 &        17.42 &         5.88 &         0.87 &        18.25 &        27.97 &         0.65 &         1.40 &         103 &         101 &         0.17 &        13.81 &         4.75 \\ [0.01ex]
A1831 &        19.15 &        15.78 &         0.86 &        18.81 &        32.24 &         0.44 &         3.73 &          71 &          59 &         0.38 &        13.89 &         2.64 \\ [0.01ex]
A1983 &        17.05 &         4.62 &         0.74 &        18.67 &        11.51 &         0.59 &         3.28 &         117 &         119 &         0.70 &        14.33 &         3.25 \\ [0.01ex]
A1991 &        19.23 &        16.99 &         0.81 &        18.87 &        28.01 &         0.47 &         3.08 &         101 &          98 &         0.46 &        13.90 &         6.89 \\ [0.01ex]
A2107 &        18.87 &        16.26 &         0.86 &        19.36 &        28.30 &         0.52 &         2.88 &          17 &          33 &         0.62 &        13.12 &         2.62 \\ [0.01ex]
A2124 &        19.13 &        14.77 &         0.90 &        18.20 &        23.37 &         0.60 &         3.37 &          71 &          52 &         0.36 &        13.94 &         2.56 \\ [0.01ex]
A2149 &        17.27 &         2.92 &         0.87 &        19.21 &        20.99 &         0.73 &         2.08 &          26 &          35 &         0.23 &        15.33 &         2.83 \\ [0.01ex]
A2169 &        16.45 &         3.10 &         0.70 &        19.12 &        23.20 &         0.65 &         2.34 &         176 &         174 &         0.37 &        14.63 &         1.46 \\ [0.01ex]
A2256 &        17.33 &         5.91 &         0.86 &        17.77 &        16.69 &         0.83 &         1.28 &          56 &          27 &         0.28 &        13.90 &         3.21 \\ [0.01ex]
A2271 &        18.67 &         6.40 &         0.78 &        18.89 &        20.78 &         0.68 &         1.70 &          29 &          34 &         0.22 &        14.82 &         6.25 \\ [0.01ex]
A2382 &        18.16 &         3.42 &         0.77 &        18.82 &        12.41 &         1.00 &         1.90 &           8 &         180 &         0.26 &        14.98 &         0.35 \\ [0.01ex]
A2399 &        17.86 &         4.13 &         0.67 &        18.76 &        10.81 &         0.77 &         1.51 &         103 &         110 &         0.43 &        15.02 &         0.36 \\ [0.01ex]
A2415 &        17.25 &         4.62 &         0.80 &        18.32 &        15.26 &         0.61 &         2.84 &         121 &         114 &         0.43 &        14.57 &         2.62 \\ [0.01ex]
A2457 &        18.88 &        16.26 &         0.72 &        20.16 &        54.68 &         0.39 &         3.72 &         175 &         173 &         0.50 &        13.77 &         1.10 \\ [0.01ex]
A2572a &        16.74 &         2.52 &         0.92 &        18.06 &        24.35 &         0.65 &         2.69 &         136 &          87 &         0.10 &        13.03 &        12.16 \\ [0.01ex]
A2589 &        20.10 &        25.46 &         0.86 &        18.67 &        36.76 &         0.37 &         5.74 &          84 &          95 &         0.36 &        12.78 &         6.59 \\ [0.01ex]
A2593 &        19.79 &        13.18 &         0.73 &        17.23 &        12.55 &         0.60 &         6.18 &         168 &         164 &         0.32 &        13.53 &         2.88 \\ [0.01ex]
A2622 &        17.82 &         6.91 &         0.86 &        18.42 &        19.13 &         0.57 &         2.38 &          35 &          37 &         0.39 &        14.40 &         3.37 \\ [0.01ex]
A2657 &        18.48 &         4.74 &         0.81 &        18.12 &        16.59 &         0.60 &         2.91 &          29 &           5 &         0.15 &        14.01 &         2.11 \\ [0.01ex]
A2665 &        19.05 &        17.64 &         0.86 &        19.78 &        43.81 &         0.48 &         2.71 &          20 &           3 &         0.49 &        13.61 &         1.71 \\ [0.01ex]
A2717 &        18.22 &         3.37 &         0.96 &        18.92 &        18.60 &         1.00 &         2.47 &         168 &          11 &         0.15 &        13.76 &         0.85 \\ [0.01ex]
A2734 &        19.25 &         8.86 &         0.83 &        19.40 &        27.46 &         0.57 &         1.34 &          24 &          19 &         0.20 &        14.30 &         0.44 \\ [0.01ex]
A3128 &        18.09 &         4.17 &         0.80 &        19.03 &        18.69 &         0.79 &         2.78 &           2 &          61 &         0.26 &        14.36 &         0.97 \\ [0.01ex]
A3158 &        18.69 &         6.49 &         0.90 &        19.37 &        34.32 &         0.62 &         2.96 &          74 &          95 &         0.17 &        13.68 &         0.46 \\ [0.01ex]
A3266 &        19.96 &        16.47 &         0.86 &        19.60 &        54.47 &         0.37 &         4.30 &          68 &          72 &         0.20 &        13.26 &         1.16 \\ [0.01ex]
A3376 &        17.27 &         3.55 &         0.72 &        18.33 &        14.46 &         0.63 &         0.96 &          64 &          67 &         0.23 &        13.96 &         0.51 \\ [0.01ex]
A3395 &        19.53 &         9.81 &         0.62 &        19.10 &        25.79 &         0.34 &         3.04 &         124 &         126 &         0.24 &        14.13 &         0.66 \\ [0.01ex]
A3490 &        17.34 &         2.76 &         0.94 &        18.14 &        13.85 &         0.59 &         1.52 &          18 &          29 &         0.16 &        14.76 &         0.68 \\ [0.01ex]
A3497 &        17.43 &         2.62 &         0.92 &        19.01 &        12.45 &         0.71 &         1.62 &          57 &          35 &         0.31 &        15.42 &         0.47 \\ [0.01ex]
A3528a &        17.36 &         4.53 &         0.76 &        17.08 &         8.81 &         0.95 &         1.23 &          96 &         156 &         0.30 &        13.63 &         4.27 \\ [0.01ex]
A3528b &        17.64 &         4.07 &         0.83 &        18.68 &        21.39 &         0.59 &         2.26 &           1 &         176 &         0.21 &        13.80 &         0.98 \\ [0.01ex]
A3530 &        18.42 &         6.17 &         0.79 &        18.99 &        38.25 &         0.40 &         2.29 &         126 &         136 &         0.11 &        13.40 &         1.43 \\ [0.01ex]
A3532 &        17.97 &         4.00 &         0.86 &        19.04 &        26.39 &         0.72 &         2.34 &          46 &          90 &         0.15 &        13.69 &         0.42 \\ [0.01ex]
A3556 &        17.90 &         5.77 &         0.66 &        18.55 &        15.67 &         0.69 &         2.92 &         144 &         163 &         0.43 &        13.69 &         0.82 \\ [0.01ex]
A3558 &        18.16 &         5.47 &         0.93 &        18.04 &        20.21 &         0.62 &         1.24 &         180 &         159 &         0.12 &        13.13 &         0.88 \\ [0.01ex]
A3560 &        16.66 &         6.35 &         0.78 &        17.29 &        16.03 &         0.96 &         2.20 &          72 &          92 &         0.43 &        12.14 &         3.48 \\ [0.01ex]
A3667 &        19.48 &        12.89 &         0.82 &        19.82 &        40.30 &         0.44 &         2.41 &          67 &         146 &         0.28 &        13.70 &         0.91 \\ [0.01ex]
A3716 &        19.63 &        10.22 &         0.99 &        19.58 &        24.97 &         0.45 &         3.24 &          88 &          58 &         0.34 &        13.87 &         0.61 \\ [0.01ex]
A3809 &        20.11 &        14.85 &         0.87 &        21.06 &        48.80 &         0.39 &         2.84 &          89 &          84 &         0.40 &        14.50 &         0.26 \\ [0.01ex]
A3880 &        19.65 &        12.30 &         0.98 &        20.04 &        47.74 &         0.50 &         2.30 &           0 &         153 &         0.21 &        13.65 &         0.45 \\ [0.01ex]
A4059 &        19.03 &         9.27 &         0.75 &        18.99 &        27.51 &         0.56 &         1.97 &         160 &         158 &         0.22 &        13.36 &         0.36 \\ [0.01ex]
MKW3s &        17.62 &         2.50 &         0.97 &        18.52 &        15.55 &         0.55 &         1.61 &           6 &          12 &         0.12 &        14.82 &         2.75 \\ [0.01ex]
RXJ1022 &        19.03 &         9.70 &         0.80 &        19.42 &        18.29 &         0.51 &         3.72 &         164 &         159 &         0.58 &        14.87 &         2.23 \\ [0.01ex]
RXJ1740 &        19.43 &        15.48 &         0.74 &        18.85 &        17.58 &         0.39 &         3.63 &         109 &         108 &         0.61 &        14.00 &         1.53 \\ [0.01ex] 
ZwCl1261 &        17.89 &         8.19 &         0.85 &        18.48 &        29.34 &         0.52 &         2.48 &          42 &          50 &         0.28 &        13.94 &         2.52 \\ [0.01ex]
ZwCl2844 &        17.51 &         7.15 &         0.83 &        19.31 &        43.66 &         0.29 &         2.64 &          48 &          52 &         0.29 &        13.72 &         3.58 \\ [0.01ex]
ZwCl8338 &        18.50 &        13.03 &         0.82 &        19.14 &        24.68 &         0.59 &         4.22 &          89 &          49 &         0.65 &        13.45 &         3.14 \\ [0.01ex]
ZwCl8852 &        17.85 &         9.40 &         0.71 &        18.82 &        37.38 &         0.48 &         2.32 &         100 &         111 &         0.30 &        12.91 &         2.51 \\ [0.01ex]
A3562 &        17.49 &         7.86 &         0.78 &        18.70 &        43.30 &         0.43 &         1.46 &          89 &          81 &         0.18 &        12.46 &         0.83 \\ [0.01ex]
\hline
%\label{tab:dcat}
\end{longtable}

\newpage

\subsection[]{ACS BCGs sample}

\small
\begin{longtable}{lcccccccccccc}
            \hline
\multicolumn{1}{c}{\rm Name}&
\multicolumn{1}{c}{$\mu_e$}&
\multicolumn{1}{c}{$r_e$}&
\multicolumn{1}{c}{$e_b$}&
\multicolumn{1}{c}{$\mu_0$}&
\multicolumn{1}{c}{$h$}&
\multicolumn{1}{c}{$e_d$}&
\multicolumn{1}{c}{$n$}&
\multicolumn{1}{c}{$PA_b$}&
\multicolumn{1}{c}{$PA_d$}&
\multicolumn{1}{c}{$B/T$}&
\multicolumn{1}{c}{$m_V$}&
\multicolumn{1}{c}{$\chi^2$}\\
\multicolumn{1}{c}{}&
\multicolumn{1}{c}{mag/arc sec$^2$}&
\multicolumn{1}{c}{kpc}&
\multicolumn{1}{c}{}&
\multicolumn{1}{c}{mag/arc sec$^2$}&
\multicolumn{1}{c}{kpc}&
\multicolumn{1}{c}{}&
\multicolumn{1}{c}{}&
\multicolumn{1}{c}{}&
\multicolumn{1}{c}{}&
\multicolumn{1}{c}{}&
\multicolumn{1}{c}{}&
\multicolumn{1}{c}{}\\
\hline\noalign{\smallskip}
rxj0056 &        18.36 &         9.52 &         0.88 &        17.55 &        16.99 &         0.55 &         3.33 &          59 &          21 &         0.33 &        18.58 &         4.51 \\ [0.01ex]
rxj0110 &        16.09 &         3.38 &         0.93 &        17.37 &        14.95 &         0.75 &         1.84 &         135 &         156 &         0.29 &        17.03 &         7.81 \\ [0.01ex]
rxj0154 &        17.25 &         6.98 &         0.70 &        17.09 &         9.47 &         0.67 &         2.46 &         129 &         157 &         0.57 &        17.69 &         7.06 \\ [0.01ex]
rxj0522 &        16.64 &         2.82 &         0.80 &        17.56 &        14.35 &         0.84 &         1.86 &          19 &         120 &         0.18 &        18.45 &         3.92 \\ [0.01ex]
rxj0841 &        19.19 &        19.21 &         0.74 &        18.45 &        35.67 &         0.41 &         3.58 &          18 &          14 &         0.33 &        16.82 &         4.16 \\ [0.01ex]
rxj0847 &        16.11 &         2.97 &         0.87 &        17.07 &        12.21 &         0.83 &         1.35 &          86 &          36 &         0.24 &        18.67 &         5.42 \\ [0.01ex]
rxj0926 &        16.16 &         2.63 &         0.77 &        16.59 &        10.32 &         0.59 &         1.41 &         137 &         137 &         0.18 &        18.63 &         7.42 \\ [0.01ex]
rxj0957 &        16.68 &         2.84 &         0.94 &        17.90 &        19.66 &         0.75 &         3.59 &           0 &          20 &         0.18 &        18.48 &         8.13 \\ [0.01ex]
rxj1117 &        16.87 &         2.75 &         0.83 &        18.05 &        10.28 &         0.68 &         1.93 &         130 &         155 &         0.35 &        18.44 &         5.42 \\ [0.01ex]
rxj1123 &        16.82 &         3.40 &         0.89 &        17.46 &        14.61 &         0.75 &         3.18 &          55 &         107 &         0.24 &        17.45 &        21.13 \\ [0.01ex]
rxj1354 &        17.13 &         8.08 &         0.79 &        17.43 &        18.02 &         0.78 &         2.92 &         116 &          97 &         0.45 &        17.83 &         8.11 \\ [0.01ex]
rxj1540 &        15.96 &         2.07 &         0.94 &        16.85 &         7.35 &         0.97 &         1.89 &         162 &          99 &         0.31 &        18.48 &         1.37 \\ [0.01ex]
rxj1642 &        16.65 &         1.73 &         0.99 &        17.71 &        10.81 &         0.88 &         1.63 &         142 &          77 &         0.14 &        18.47 &         8.30 \\ [0.01ex]
rxj2059 &        17.48 &         2.68 &         0.89 &        18.32 &        14.90 &         0.89 &         2.84 &          70 &          89 &         0.18 &        18.09 &         5.14 \\ [0.01ex]
rxj2108 &        17.57 &         7.32 &         0.83 &        18.64 &        24.17 &         0.59 &         2.51 &          91 &          77 &         0.42 &        17.23 &         6.15 \\ [0.01ex]
rxj2139 &        16.40 &         5.09 &         0.60 &        17.49 &        13.34 &         0.49 &         2.92 &         164 &         170 &         0.55 &        17.75 &         5.12 \\ [0.01ex]
rxj2202 &        15.67 &         2.80 &         0.82 &        17.44 &        13.45 &         0.89 &         2.10 &          33 &          75 &         0.37 &        17.96 &         6.57 \\ [0.01ex]
rxj2328 &        16.09 &         2.73 &         0.95 &        17.39 &        15.16 &         0.77 &         1.94 &          17 &          83 &         0.21 &        18.28 &         6.08 \\ [0.01ex]
rxj0826 &        16.25 &         2.36 &         0.75 &        17.61 &        10.56 &         0.66 &         2.66 &          60 &          61 &         0.34 &        18.38 &        15.85 \\ [0.01ex]
rxj1015 &        18.05 &         5.74 &         0.86 &        18.49 &        17.14 &         0.60 &         3.83 &         113 &          86 &         0.37 &        18.36 &         4.79 \\ [0.01ex]
\hline
%\label{tab:dcat}
\end{longtable}
\begin{minipage}{175mm}
NOTE. Col. (1): Galaxy Cluster; Col. (2): Effective surface brightness of the bulge at $r_e$; Col. (3): Effective radius of the bulge;  Col. (4): Ellipticity of the bulge; Col. (5): Central surface brightness of the disk; Col. (6): Scale length of the disk; Col. (7): Ellipticity of the disk; Col. (8): Shape parameter of the bulge; Col. (9): Position angle of the bulge; Col. (10): Position angle of the disk; Col. (11): bulge-to-total luminosity ratio; Col. (12): V band rest frame magnitude calculated from the model; Col. (13): $\chi^2$ of the fit \end{minipage}

\section[]{Results of the 1-component S\'ersic fit}

\subsection[]{WINGS BCGs sample}

\small
\begin{longtable}{lcccccccccccc}
            \hline
\multicolumn{1}{c}{\rm Name}&
\multicolumn{1}{c}{$\mu_e$}&
\multicolumn{1}{c}{$r_e$}&
\multicolumn{1}{c}{$e_b$}&
\multicolumn{1}{c}{$n$}&
\multicolumn{1}{c}{$PA_b$}&
\multicolumn{1}{c}{$m_V$}&
\multicolumn{1}{c}{$\chi^2$}\\
\multicolumn{1}{c}{}&
\multicolumn{1}{c}{mag/arc sec$^2$}&
\multicolumn{1}{c}{kpc}&
\multicolumn{1}{c}{}&
\multicolumn{1}{c}{}&
\multicolumn{1}{c}{}&
\multicolumn{1}{c}{}\\
\hline\noalign{\smallskip}
A85 &        18.63 &        25.07 &         0.73 &         1.51 &          57 &        13.39 &         3.97 \\ [0.01ex]
A119 &        20.92 &        64.93 &         0.78 &         4.92 &         126 &        12.47 &         4.85 \\ [0.01ex]
A147 &        18.50 &        11.95 &         0.76 &         2.69 &         148 &        14.10 &         4.28 \\ [0.01ex]
A151 &        20.61 &        52.84 &         0.78 &         4.97 &         152 &        13.00 &         8.67 \\ [0.01ex]
A168 &        19.47 &        20.14 &         0.86 &         4.66 &          61 &        13.52 &         3.13 \\ [0.01ex]
A311 &        20.06 &        36.82 &         0.68 &         4.00 &         115 &        13.92 &        29.69 \\ [0.01ex]
A376 &        21.25 &        49.33 &         0.90 &         5.99 &           4 &        13.30 &         4.57 \\ [0.01ex]
A500 &        21.12 &        31.48 &         0.77 &         5.74 &          11 &        14.34 &         0.78 \\ [0.01ex]
A548b &        21.74 &        38.02 &         0.95 &         4.69 &         168 &        13.43 &         1.15 \\ [0.01ex]
A602 &        20.74 &        31.50 &         0.87 &         3.57 &          67 &        14.60 &        10.35 \\ [0.01ex]
A671 &        20.68 &        56.78 &         0.76 &         5.60 &         117 &        12.77 &         6.52 \\ [0.01ex]
A754 &        20.28 &        47.61 &         0.69 &         4.58 &          20 &        13.11 &         5.39 \\ [0.01ex]
A957 &        20.86 &        59.30 &         0.81 &         5.70 &           0 &        12.54 &         7.89 \\ [0.01ex]
A970 &        21.37 &        37.07 &         0.88 &         6.70 &          32 &        14.44 &         4.65 \\ [0.01ex]
A1069 &        20.45 &        39.89 &         0.77 &         5.30 &           0 &        13.83 &        12.32 \\ [0.01ex]
A1291 &        21.05 &        37.74 &         0.69 &         4.50 &         170 &        14.24 &         7.41 \\ [0.01ex]
A1631a &        21.43 &        38.65 &         0.74 &         6.59 &         146 &        13.40 &         0.87 \\ [0.01ex]
A1644 &        21.75 &        77.32 &         0.68 &         3.84 &          45 &        12.62 &         3.97 \\ [0.01ex]
A1668 &        20.25 &        28.16 &         0.79 &         3.45 &         161 &        14.53 &         2.66 \\ [0.01ex]
A1736 &        19.71 &        23.51 &         0.60 &         4.21 &         136 &        13.20 &         1.38 \\ [0.01ex]
A1795 &        21.57 &        94.66 &         0.74 &         5.04 &          99 &        13.09 &        97.00 \\ [0.01ex]
A1831 &        21.56 &        81.93 &         0.77 &         5.88 &          64 &        13.27 &         3.99 \\ [0.01ex]
A1983 &        17.91 &         7.89 &         0.72 &         4.17 &         118 &        14.25 &         3.61 \\ [0.01ex]
A1991 &        20.16 &        39.01 &         0.72 &         3.66 &          99 &        13.64 &         7.62 \\ [0.01ex]
A2107 &        19.46 &        26.26 &         0.81 &         3.31 &          23 &        13.00 &         2.89 \\ [0.01ex]
A2124 &        20.86 &        62.26 &         0.80 &         4.48 &          58 &        13.37 &         4.28 \\ [0.01ex]
A2149 &        20.61 &        16.91 &         0.87 &         7.90 &          29 &        15.59 &        10.69 \\ [0.01ex]
A2169 &        18.88 &        11.44 &         0.69 &         5.48 &         175 &        14.83 &         4.83 \\ [0.01ex]
A2256 &        18.98 &        20.62 &         0.86 &         2.63 &          42 &        13.81 &         5.82 \\ [0.01ex]
A2271 &        22.33 &        79.69 &         0.75 &         5.04 &          32 &        14.03 &         4.35 \\ [0.01ex]
A2382 &        22.32 &        50.20 &         0.87 &         5.50 &          10 &        14.30 &         0.59 \\ [0.01ex]
A2399 &        19.31 &        10.96 &         0.69 &         2.80 &         104 &        14.99 &         0.71 \\ [0.01ex]
A2415 &        19.60 &        19.70 &         0.75 &         5.31 &         118 &        14.28 &         3.41 \\ [0.01ex]
A2457 &        20.16 &        36.09 &         0.70 &         5.00 &         174 &        13.67 &         1.33 \\ [0.01ex]
A2572a &        21.33 &        49.27 &         0.86 &         6.30 &         121 &        12.99 &        42.62 \\ [0.01ex]
A2589 &        21.61 &        85.79 &         0.72 &         6.10 &          92 &        12.45 &         8.63 \\ [0.01ex]
A2593 &        20.91 &        56.56 &         0.65 &         5.21 &         164 &        12.81 &        11.15 \\ [0.01ex]
A2622 &        19.29 &        19.87 &         0.79 &         3.63 &          36 &        14.23 &         4.00 \\ [0.01ex]
A2657 &        21.18 &        44.40 &         0.73 &         4.30 &          15 &        13.53 &         3.41 \\ [0.01ex]
A2665 &        19.90 &        33.05 &         0.83 &         3.38 &          15 &        13.53 &         1.93 \\ [0.01ex]
A2717 &        21.32 &        31.30 &         1.00 &         4.30 &         180 &        13.77 &         2.78 \\ [0.01ex]
A2734 &        20.94 &        35.89 &         0.75 &         2.70 &          22 &        14.14 &         0.50 \\ [0.01ex]
A3128 &        20.84 &        24.47 &         0.87 &         4.90 &          16 &        14.32 &         1.81 \\ [0.01ex]
A3158 &        22.03 &        59.79 &         0.86 &         5.70 &          84 &        13.49 &         1.04 \\ [0.01ex]
A3266 &        21.80 &        63.21 &         0.73 &         5.90 &          74 &        13.28 &         2.32 \\ [0.01ex]
A3376 &        19.36 &        14.66 &         0.68 &         2.75 &          66 &        13.98 &         2.01 \\ [0.01ex]
A3395 &        21.46 &        46.49 &         0.54 &         4.30 &         125 &        13.75 &         1.01 \\ [0.01ex]
A3490 &        21.09 &        32.11 &         0.87 &         5.01 &           0 &        14.25 &         2.07 \\ [0.01ex]
A3497 &        19.37 &         8.77 &         0.88 &         3.48 &          47 &        15.49 &         1.28 \\ [0.01ex]
A3528a &        18.44 &        12.50 &         0.90 &         2.09 &         104 &        13.55 &         4.70 \\ [0.01ex]
A3528b &        21.94 &        51.49 &         0.77 &         7.90 &         180 &        13.44 &         2.85 \\ [0.01ex]
A3530 &        21.93 &        71.45 &         0.70 &         5.10 &         133 &        13.09 &         2.11 \\ [0.01ex]
A3532 &        21.93 &        39.08 &         0.95 &         7.10 &          36 &        13.95 &         4.54 \\ [0.01ex]
A3556 &        20.77 &        34.83 &         0.67 &         5.61 &         147 &        13.23 &         1.18 \\ [0.01ex]
A3558 &        20.86 &        52.57 &         0.77 &         3.30 &         161 &        12.56 &         1.28 \\ [0.01ex]
A3560 &        18.49 &        19.93 &         1.00 &         3.67 &         126 &        12.01 &         7.62 \\ [0.01ex]
A3667 &        21.15 &        39.25 &         0.93 &         3.76 &          88 &        13.51 &         1.27 \\ [0.01ex]
A3716 &        21.34 &        33.17 &         1.00 &         4.64 &           0 &        13.50 &         1.11 \\ [0.01ex]
A3809 &        21.03 &        28.28 &         0.83 &         3.61 &          87 &        14.49 &         0.29 \\ [0.01ex]
A3880 &        21.23 &        41.88 &         0.90 &         3.44 &         159 &        13.63 &         0.72 \\ [0.01ex]
A4059 &        21.17 &        51.12 &         0.68 &         3.54 &         159 &        13.00 &         0.49 \\ [0.01ex]
MKW3s &        22.35 &        58.11 &         0.75 &         5.90 &          12 &        14.09 &         4.41 \\ [0.01ex]
RXJ1022 &        20.23 &        22.13 &         0.76 &         4.78 &         163 &        14.62 &         2.52 \\ [0.01ex]
RXJ1740 &        19.05 &        16.82 &         0.65 &         2.90 &         110 &        14.02 &         4.40 \\ [0.01ex]
ZwCl1261 &        20.40 &        44.64 &         0.77 &         4.70 &          46 &        13.58 &         3.97 \\ [0.01ex]
ZwCl2844 &        18.87 &        16.87 &         0.74 &         4.05 &          50 &        13.77 &         6.44 \\ [0.01ex]
ZwCl8338 &        21.12 &        52.44 &         0.90 &         7.90 &         108 &        12.97 &        10.23 \\ [0.01ex]
ZwCl8852 &        20.36 &        47.84 &         0.66 &         4.61 &         104 &        12.63 &         3.63 \\ [0.01ex]
A3562 &        19.01 &        23.45 &         0.70 &         2.68 &          85 &        12.76 &         1.37 \\ [0.01ex]
\hline
%\label{tab:dcat}
\end{longtable}

\subsection[]{ACS BCGs sample}

\small
\begin{longtable}{lcccccccccccc}
            \hline
\multicolumn{1}{c}{\rm Name}&
\multicolumn{1}{c}{$\mu_e$}&
\multicolumn{1}{c}{$r_e$}&
\multicolumn{1}{c}{$e_b$}&
\multicolumn{1}{c}{$n$}&
\multicolumn{1}{c}{$PA_b$}&
\multicolumn{1}{c}{$m_V$}&
\multicolumn{1}{c}{$\chi^2$}\\
\multicolumn{1}{c}{}&
\multicolumn{1}{c}{mag/arc sec$^2$}&
\multicolumn{1}{c}{kpc}&
\multicolumn{1}{c}{}&
\multicolumn{1}{c}{}&
\multicolumn{1}{c}{}&
\multicolumn{1}{c}{}\\
\hline\noalign{\smallskip}
rxj0056 &        18.57 &        13.90 &         0.90 &         3.10 &          59  &        18.87 &         3.26 \\ [0.01ex]
rxj0110 &        18.70 &        17.41 &         0.88 &         4.25 &         146  &        16.88 &        11.16 \\ [0.01ex]
rxj0154 &        17.86 &        12.36 &         0.72 &         2.80 &         136  &        17.55 &         7.91 \\ [0.01ex]
rxj0522 &        20.89 &        34.66 &         0.90 &         6.70 &          12  &        18.34 &         4.41 \\ [0.01ex]
rxj0841 &        21.08 &        79.17 &         0.69 &         4.90 &          17  &        16.36 &         1.27 \\ [0.01ex]
rxj0847 &        18.77 &        17.11 &         0.90 &         3.59 &          68  &        18.52 &         8.50 \\ [0.01ex]
rxj0926 &        18.84 &        19.49 &         0.66 &         3.56 &         137  &        18.30 &         8.68 \\ [0.01ex]
rxj0957 &        20.38 &        41.96 &         0.30 &         7.50 &         168  &        18.84 &        41.41 \\ [0.01ex]
rxj1117 &        18.63 &         8.61 &         0.81 &         3.51 &         138  &        18.43 &         7.80 \\ [0.01ex]
rxj1123 &        20.29 &        25.11 &         0.96 &         7.90 &          72  &        17.43 &        11.76 \\ [0.01ex]
rxj1354 &        19.14 &        31.77 &         0.79 &         4.61 &         115  &        17.49 &         9.79 \\ [0.01ex]
rxj1540 &        19.26 &        16.52 &         0.96 &         4.98 &         156  &        18.23 &         6.10 \\ [0.01ex]
rxj1642 &        20.95 &        22.46 &         1.00 &         6.70 &           0  &        18.49 &         9.54 \\ [0.01ex]
rxj2059 &        21.42 &        25.84 &         0.97 &         7.70 &          62  &        18.36 &         6.95 \\ [0.01ex]
rxj2108 &        18.99 &        18.86 &         0.80 &         3.75 &          86  &        17.18 &         6.96 \\ [0.01ex]
rxj2139 &        17.78 &        12.35 &         0.58 &         4.24 &         165  &        17.60 &         5.80 \\ [0.01ex]
rxj2202 &        17.39 &         7.74 &         0.86 &         3.70 &          32  &        18.27 &         4.42 \\ [0.01ex]
rxj2328 &        19.86 &        21.15 &         1.00 &         7.10 &          26  &        18.37 &        17.98 \\ [0.01ex]
rxj0826 &        19.08 &        11.71 &         0.76 &         6.30 &          60  &        18.35 &        14.13 \\ [0.01ex]
rxj1015 &        20.95 &        31.36 &         0.87 &         7.30 &         113  &        18.07 &         2.59 \\ [0.01ex]
\hline
%\label{tab:dcat}
\end{longtable}
\begin{minipage}{175mm}
NOTE. Col. (1): Galaxy Cluster; Col. (2): Effective surface brightness of the bulge at $r_e$; Col. (3): Effective radius of the bulge;  Col. (4): Ellipticity of the bulge; Col. (5): Shape parameter of the bulge; Col. (6): Position angle of the bulge; Col. (7): V band rest frame magnitude calculated from the model; Col. (8): $\chi^2$ of the fit \end{minipage}

\section[]{Results of the 1-component de Vaucouleur fit}

\subsection[]{WINGS BCGs sample}

\small
\begin{longtable}{lcccccc}
            \hline
\multicolumn{1}{c}{\rm Name}&
\multicolumn{1}{c}{$\mu_e$}&
\multicolumn{1}{c}{$r_e$}&
\multicolumn{1}{c}{$e_b$}&
\multicolumn{1}{c}{$PA_b$}&
\multicolumn{1}{c}{$m_V$}&
\multicolumn{1}{c}{$\chi^2$}\\
\multicolumn{1}{c}{}&
\multicolumn{1}{c}{mag/arc sec$^2$}&
\multicolumn{1}{c}{kpc}&
\multicolumn{1}{c}{}&
\multicolumn{1}{c}{}&
\multicolumn{1}{c}{}&
\multicolumn{1}{c}{}\\
\hline\noalign{\smallskip}
A85 &        21.33 &       112.55 &         0.73 &          57 &        12.32 &        20.12 \\ [0.01ex]
A119 &        20.09 &        40.90 &         0.78 &         126 &        12.76 &         5.46 \\ [0.01ex]
A147 &        19.68 &        22.87 &         0.77 &         149 &        13.66 &        16.40 \\ [0.01ex]
A151 &        19.82 &        34.24 &         0.78 &         151 &        13.26 &         9.25 \\ [0.01ex]
A168 &        19.02 &        16.04 &         0.86 &          61 &        13.64 &         3.54 \\ [0.01ex]
A311 &        20.06 &        36.86 &         0.68 &         115 &        13.92 &        29.81 \\ [0.01ex]
A376 &        19.62 &        20.01 &         0.90 &           3 &        13.83 &         7.06 \\ [0.01ex]
A500 &        19.82 &        15.86 &         0.77 &          11 &        14.72 &         1.13 \\ [0.01ex]
A548b &        21.06 &        25.47 &         0.96 &         168 &        13.70 &         1.24 \\ [0.01ex]
A602 &        21.15 &        39.89 &         0.87 &          67 &        14.44 &        10.50 \\ [0.01ex]
A671 &        19.53 &        30.92 &         0.76 &         116 &        13.11 &         8.18 \\ [0.01ex]
A754 &        19.81 &        36.97 &         0.69 &          20 &        13.26 &         5.83 \\ [0.01ex]
A957 &        19.58 &        30.23 &         0.81 &         178 &        12.92 &        19.73 \\ [0.01ex]
A970 &        19.37 &        12.50 &         0.91 &          32 &        15.04 &         7.36 \\ [0.01ex]
A1069 &        19.77 &        30.42 &         0.72 &         180 &        13.96 &        30.53 \\ [0.01ex]
A1291 &        20.68 &        31.30 &         0.69 &         169 &        14.35 &         8.04 \\ [0.01ex]
A1631a &        19.52 &        14.08 &         0.73 &         146 &        13.96 &         2.04 \\ [0.01ex]
A1644 &        21.92 &        86.20 &         0.68 &          45 &        12.54 &         3.98 \\ [0.01ex]
A1668 &        20.67 &        35.07 &         0.79 &         161 &        14.40 &         2.88 \\ [0.01ex]
A1736 &        19.54 &        21.47 &         0.60 &         136 &        13.25 &         1.39 \\ [0.01ex]
A1795 &        20.66 &        57.30 &         0.74 &         100 &        13.39 &        97.26 \\ [0.01ex]
A1831 &        20.02 &        34.99 &         0.77 &          65 &        13.77 &         4.83 \\ [0.01ex]
A1983 &        17.82 &         7.55 &         0.72 &         118 &        14.27 &         3.67 \\ [0.01ex]
A1991 &        20.43 &        45.07 &         0.72 &          99 &        13.56 &         7.70 \\ [0.01ex]
A2107 &        20.00 &        34.72 &         0.81 &          23 &        12.83 &         3.40 \\ [0.01ex]
A2124 &        20.44 &        49.50 &         0.80 &          58 &        13.51 &         4.37 \\ [0.01ex]
A2149 &        19.57 &        12.18 &         0.84 &          27 &        15.66 &         7.51 \\ [0.01ex]
A2169 &        17.95 &         7.15 &         0.70 &         175 &        15.07 &         5.72 \\ [0.01ex]
A2256 &        20.19 &        39.28 &         0.87 &          42 &        13.40 &        12.22 \\ [0.01ex]
A2271 &        21.43 &        48.16 &         0.75 &          32 &        14.34 &         4.59 \\ [0.01ex]
A2382 &        21.03 &        24.29 &         0.88 &          10 &        14.75 &         0.87 \\ [0.01ex]
A2399 &        20.27 &        18.14 &         0.69 &         104 &        14.69 &         1.24 \\ [0.01ex]
A2415 &        18.81 &        13.34 &         0.76 &         116 &        14.48 &         5.02 \\ [0.01ex]
A2457 &        19.40 &        24.11 &         0.70 &         174 &        13.90 &         1.75 \\ [0.01ex]
A2572a &        20.37 &        36.49 &         0.80 &         106 &        13.00 &        22.63 \\ [0.01ex]
A2589 &        20.05 &        39.70 &         0.72 &          92 &        12.77 &        21.86 \\ [0.01ex]
A2593 &        20.01 &        34.98 &         0.65 &         164 &        13.08 &        11.75 \\ [0.01ex]
A2622 &        19.59 &        23.27 &         0.79 &          36 &        14.14 &         4.15 \\ [0.01ex]
A2657 &        20.98 &        40.62 &         0.73 &          12 &        13.56 &         4.09 \\ [0.01ex]
A2665 &        20.46 &        45.15 &         0.83 &          15 &        13.33 &         2.22 \\ [0.01ex]
A2717 &        21.12 &        28.80 &         0.97 &         121 &        13.81 &         2.91 \\ [0.01ex]
A2734 &        22.34 &        82.28 &         0.74 &          21 &        13.54 &         0.94 \\ [0.01ex]
A3128 &        20.24 &        18.15 &         0.87 &          14 &        14.48 &         2.42 \\ [0.01ex]
A3158 &        20.65 &        28.19 &         0.85 &          86 &        13.93 &         2.05 \\ [0.01ex]
A3266 &        20.64 &        38.31 &         0.70 &          72 &        13.47 &         3.00 \\ [0.01ex]
A3376 &        20.51 &        27.44 &         0.68 &          66 &        13.59 &         2.69 \\ [0.01ex]
A3395 &        21.23 &        42.05 &         0.53 &         125 &        13.81 &         1.12 \\ [0.01ex]
A3490 &        20.25 &        20.21 &         0.87 &           0 &        14.52 &         2.22 \\ [0.01ex]
A3497 &        19.79 &        10.94 &         0.88 &          47 &        15.36 &         1.36 \\ [0.01ex]
A3528a &        20.02 &        25.44 &         1.00 &         180 &        13.13 &       \\ [0.01ex]
A3528b &        20.16 &        23.52 &         0.74 &         179 &        13.76 &         1.86 \\ [0.01ex]
A3530 &        20.93 &        40.76 &         0.68 &         132 &        13.46 &         2.79 \\ [0.01ex]
A3532 &        20.15 &        17.52 &         0.87 &          40 &        14.30 &         2.45 \\ [0.01ex]
A3556 &        19.50 &        17.47 &         0.68 &         147 &        13.63 &         1.83 \\ [0.01ex]
A3558 &        21.61 &        83.06 &         0.76 &         159 &        12.23 &         2.33 \\ [0.01ex]
A3560 &        18.78 &        25.60 &         0.83 &          73 &        11.90 &         4.36 \\ [0.01ex]
A3667 &        21.37 &        44.67 &         0.93 &          88 &        13.43 &         1.27 \\ [0.01ex]
A3716 &        20.78 &        24.40 &         1.00 &          64 &        13.69 &         1.16 \\ [0.01ex]
A3809 &        21.38 &        34.45 &         0.83 &          87 &        14.37 &         0.31 \\ [0.01ex]
A3880 &        21.80 &        58.25 &         0.90 &         159 &        13.41 &         0.75 \\ [0.01ex]
A4059 &        21.62 &        66.61 &         0.68 &         159 &        12.82 &         0.53 \\ [0.01ex]
MKW3s &        20.77 &        24.70 &         0.74 &          12 &        14.59 &         5.10 \\ [0.01ex]
RXJ1022 &        19.74 &        17.42 &         0.75 &         164 &        14.75 &         2.88 \\ [0.01ex]
RXJ1740 &        19.94 &        26.02 &         0.67 &         109 &        13.77 &         3.12 \\ [0.01ex]
ZwCl1261 &        19.81 &        32.21 &         0.77 &          46 &        13.78 &         4.35 \\ [0.01ex]
ZwCl2844 &        18.84 &        16.57 &         0.74 &          50 &        13.78 &         6.44 \\ [0.01ex]
ZwCl8338 &        18.77 &        17.30 &         0.84 &          79 &        13.47 &         6.75 \\ [0.01ex]
ZwCl8852 &        19.82 &        35.50 &         0.66 &         104 &        12.81 &         3.61 \\ [0.01ex]
A3562 &        20.30 &        48.54 &         0.69 &          85 &        12.28 &         1.72 \\ [0.01ex]
\hline
%\label{tab:dcat}
\end{longtable}

\subsection[]{ACS BCGs sample}

\small
\begin{longtable}{lcccccc}
            \hline
\multicolumn{1}{c}{\rm Name}&
\multicolumn{1}{c}{$\mu_e$}&
\multicolumn{1}{c}{$r_e$}&
\multicolumn{1}{c}{$e_b$}&
\multicolumn{1}{c}{$PA_b$}&
\multicolumn{1}{c}{$m_V$}&
\multicolumn{1}{c}{$\chi^2$}\\
\multicolumn{1}{c}{}&
\multicolumn{1}{c}{mag/arc sec$^2$}&
\multicolumn{1}{c}{kpc}&
\multicolumn{1}{c}{}&
\multicolumn{1}{c}{}&
\multicolumn{1}{c}{}&
\multicolumn{1}{c}{}\\
\hline\noalign{\smallskip}
rxj0056 &        19.64 &	28.82 & 	0.85 &  	41  &	     18.27 &	     5.37 \\ [0.01ex]
rxj0110 &        18.51 &	15.75 & 	0.88 &         146  &	     16.94 &	    11.25 \\ [0.01ex]
rxj0154 &        18.79 &	19.98 & 	0.72 &         136  &	     17.26 &	    11.51 \\ [0.01ex]
rxj0522 &       19.80 &	23.89 & 	0.92 &  	 6  &	     18.31 &	     7.56 \\ [0.01ex]
rxj0841 &        20.34 &	56.45 & 	0.65 &  	16  &	     16.52 &	     5.26 \\ [0.01ex]
rxj0847 &        19.14 &	20.97 & 	0.90 &  	68  &	     18.39 &	     8.64 \\ [0.01ex]
rxj0926 &        19.22 &	23.92 & 	0.66 &         137  &	     18.17 &	     8.85 \\ [0.01ex]
rxj0957 &        18.73 &	13.10 & 	0.89 &  	 0  &	     18.88 &	    18.56 \\ [0.01ex]
rxj1117 &        19.03 &	10.59 & 	0.81 &         138  &	     18.31 &	     8.21 \\ [0.01ex]
rxj1123 &        19.13 &	18.51 & 	0.89 &  	87  &	     17.38 &	    29.49 \\ [0.01ex]
rxj1354 &        18.57 &	22.88 & 	0.79 &         116  &	     17.71 &	     9.96 \\ [0.01ex]
rxj1540 &        18.44 &	10.47 & 	0.97 &         151  &	     18.51 &	     7.53 \\ [0.01ex]
rxj1642 &        20.18 &	20.06 & 	1.00 &  	 0  &	     18.24 &	    12.12 \\ [0.01ex]
rxj2059 &        20.25 &	19.13 & 	1.00 &  	48  &	     18.16 &	    11.35 \\ [0.01ex]
rxj2108 &        19.19 &	21.03 & 	0.80 &  	86  &	     17.11 &	     7.03 \\ [0.01ex]
rxj2139 &        17.63 &	11.46 & 	0.58 &         166  &	     17.64 &	     6.05 \\ [0.01ex]
rxj2202 &        17.74 &	 9.58 & 	0.86 &  	36  &	     18.11 &	    12.52 \\ [0.01ex]
rxj2328 &        18.73 &	15.24 & 	0.94 &  	69  &	     18.32 &	    11.03 \\ [0.01ex]
rxj0826 &        18.14 &	 8.19 & 	0.72 &  	61  &	     18.47 &	    24.45 \\ [0.01ex]
rxj1015 &        19.04 &	12.85 & 	0.84 &         103  &	     18.46 &	     8.04 \\ [0.01ex]
\hline
%\label{tab:dcat}
\end{longtable}
\begin{minipage}{175mm}
NOTE. Col. (1): Galaxy Cluster; Col. (2): Effective surface brightness of the bulge at $r_e$; Col. (3): Effective radius of the bulge;  Col. (4): Ellipticity of the bulge; Col. (5): Position angle of the bulge; Col. (6): V band rest frame magnitude calculated from the model; Col. (7): $\chi^2$ of the fit \end{minipage}


\begin{thebibliography}{}
\bibitem[Abadi et al.(2006)]{abadi06} Abadi, M.~G., Navarro, J.~F., \& Steinmetz, M.\ 2006, \mnras, 365, 747 
\bibitem[Aguerri, Balcells \& Peletier(2001)]{aguerri01} Aguerri, J.~A.~L., Balcells, M., \& Peletier, R.~F.\ 2001, \aap, 367, 428 
\bibitem[Aguerri et al. (2004)]{aguerri04} Aguerri, J.A.L., Iglesias-P\'aramo, J., Vilchez J.M et al., 2004, \apj, 127, 1344
\bibitem[Aguerri et al.(2005a)]{aguerri05} Aguerri, J.~A.~L., Iglesias-P{\'a}ramo, J., V{\'{\i}}lchez, J.~M. et al., 2005, \aj, 130, 475 
\bibitem[Arag{\'o}n-Salamanca et al.(1993)]{aragon-salamanca93} Arag{\'o}n-Salamanca, A., Ellis, R.~S., Couch, W.~J. et al.\ 1993, \mnras, 262, 764 
\bibitem[Arag{\'o}n-Salamanca et al.(1998)]{aragon-salamanca98} Arag{\'o}n-Salamanca, A., Baugh, C.~M., \& Kauffmann, G.\ 1998, \mnras, 297, 427 
\bibitem[Arnaboldi et al.(2002)]{arnaboldi02} Arnaboldi, M., et al.\ 2002, \aj, 123, 760 
\bibitem[Arnaboldi et al.(2004)]{arnaboldi04} Arnaboldi, M., Gerhard, O., Aguerri, J.~A.~L. et al.\ 2004, \apjl, 614, L33
\bibitem[Ascaso et al.(2008)]{ascaso08} Ascaso, B., Moles, M., Aguerri, J.~A.~L. et al.\ 2008, \aap, 487, 453
\bibitem[Ascaso(2008)]{ascaso08b} Ascaso B., 2008, Tesis Doctoral, Universidad de Granada.
\bibitem[Ascaso et  al.(2009)]{ascaso09} Ascaso, B., Aguerri, J.~A.~L., Moles, M. et al.\ 2009, \aap, 506, 1071 
\bibitem[Bernardi et al.(2007)]{bernardi07} Bernardi, M., Hyde, J.~B., Sheth, R.~K. et al.\ 2007, \aj, 133, 1741 
\bibitem[Bernardi(2009)]{bernardi09} Bernardi, M.\ 2009, \mnras, 395, 1491
\bibitem[Bildfell et al.(2008)]{bildfell08} Bildfell, C., Hoekstra, H., Babul, A. et al.\ 2008, \mnras, 389, 1637
\bibitem[Bower, Lucey \& Ellis(1992a)]{bower92a} Bower, R.~G., Lucey, J.~R., \& Ellis, R.~S.\ 1992, \mnras, 254, 589 
\bibitem[Burke et al.(2000)]{burke00} Burke, D.~J., Collins, C.~A., \& Mann, R.~G.\ 2000, \apjl, 532, L105 
\bibitem[Brough et al.(2002)]{brough02} Brough, S., Collins, C.~A., Burke, D.~J. et al.\ 2002, \mnras, 329, L53
\bibitem[Caon, Capaccioli \& D'Onofrio(1993)]{caon93} Caon, N., Capaccioli, M., \& D'Onofrio, M.\ 1993, \mnras, 265, 1013 
\bibitem[Castro-Rodr\'iguez et al.(2009)]{castro-rodriguez09} Castro-Rodr{\'i}guez, N., Arnaboldi, M., Aguerri, J.~A.~L. et al.\ 2009, \aap, 507, 621 
\bibitem[Cava et al.(2009)]{cava09} Cava, A., et al.\ 2009, \aap, 495, 707
\bibitem[Collins et al.(2003)]{collins03} Collins, C., Brough, S., Burke, D. et al.\ 2003, \apss, 285, 51
\bibitem[Collins et al.(2009)]{collins09} Collins, C.~A., Stott, J. P., Hilton, M.  et al.\ 2009, \nat, 458, 603 
\bibitem[Coziol et al.(2009)]{coziol09} Coziol, R., Andernach, H., Caretta, C.~A. et al.\ 2009, \aj, 137, 4795 
\bibitem[Daddi et al.(2005)]{daddi05} Daddi, E., Renzini, A., Pirzkal, N., et al.\ 2005, \apj, 626, 680 
\bibitem[De Lucia \& Blaizot(2007)]{deLucia07b} De Lucia, G., \& Blaizot, J.\ 2007, \mnras, 375, 2 
\bibitem[Doherty et al.(2009)]{doherty09} Doherty, M., et al.\ 2009, \aap, 502, 771
\bibitem[D'Onofrio(2001)]{donofrio01} D'Onofrio, M.\ 2001, \mnras, 326, 1517
\bibitem[Dubinski(1998)]{dubinski98} Dubinski, J.\ 1998, \apj, 502, 141
\bibitem[Ebeling et al.(1996)]{ebeling96} Ebeling, H., Voges, W., Bohringer, H. et al.\ 1996, \mnras, 281, 799 
\bibitem[Eliche-Moral et al.(2006)]{eliche-moral06} Eliche-Moral, M.~C., Balcells, M., Aguerri, J.~A.~L. et al.\ 2006, \aap, 457, 91
\bibitem[Fabian, Nulsen \& Canizares(1982)]{fabian82} Fabian, A.~C., Nulsen, P.~E.~J., \& Canizares, C.~R.\ 1982, \mnras, 201, 933
\bibitem[Fan et al.(2008)]{fan08} Fan, L., Lapi, A., De Zotti, G. et al.\ 2008, \apjl, 689, L101
\bibitem[Fasano et al.(2006)]{fasano06} Fasano, G., Marmo, C., Varela, J, et al.\ 2006, \aap, 445, 805
\bibitem[Fasano et al.(2010)]{fasano10} Fasano, G., Bettoni, D., Ascaso, B., et al.\ 2010, \mnras, 294  
\bibitem[Finoguenov et al.(2001)]{finoguenov01} Finoguenov, A., Reiprich, T.~H., {\& Bo}hringer, H.\ 2001, \aap, 368, 749 
\bibitem[Freeman(1970)]{freeman70} Freeman, K.C., 1970, \apj, 160, 811
\bibitem[Gerhard et al.(2007)]{gerhard07} Gerhard, O., Arnaboldi, M., Freeman, K.~C. et al.\ 2007, \aap, 468, 815 
\bibitem[Gonzalez et al.(2003)]{gonzalez03} Gonzalez, A.~H., Zabludoff, A.~I., \& Zaritsky, D.\ 2003, \apss, 285, 67 
\bibitem[Gonzalez et al.(2005)]{gonzalez05} Gonzalez, A.~H., Zabludoff, A.~I., \& Zaritsky, D.\ 2005, \apj, 618, 195
\bibitem[Graham et al.(1996)]{graham96} Graham, A., Lauer, T.~R., Colless, M. et al.\ 1996, \apj, 465, 534 
\bibitem[Graham \& Guzm{\'a}n(2003)]{graham03} Graham, A.~W., \& Guzm{\'a}n, R.\ 2003, \aj, 125, 2936
\bibitem[Gunn \& Oke(1975)]{gunn75} Gunn, J.~E., \& Oke, J.~B.\ 1975, \apj, 195, 255
\bibitem[Guti{\'e}rrez et al.(2004)]{gutierrez04} Guti{\'e}rrez, C.~M., Trujillo, I., Aguerri, J.~A.~L. et al.\ 2004, \apj, 602, 664 
\bibitem[Hoessel \& Schneider(1985)]{hoessel85} Hoessel, J.~G., \& Schneider, D.~P.\ 1985, \aj, 90, 1648 
\bibitem[Hopkins et al.(2010)]{hopkins10} Hopkins, P.~F., Bundy, K., Hernquist, L. et al.\ 2010, \mnras, 401, 1099 
\bibitem[Humason, Mayall \& Sandage(1956)]{humason56} Humason, M.~L., Mayall, N.~U., \& Sandage, A.~R.\ 1956, \aj, 61, 97
\bibitem[Jones \& Forman(1984)]{jones84} Jones, C., \& Forman, W.\ 1984, \apj, 276, 38 
\bibitem[Kim et al.(2002)]{kim02} Kim, R.~S.~J., Annis, J.,  Strauss, M.~A., et al..\ 2002, Tracing Cosmic Evolution with Galaxy Clusters, 268, 395 
\bibitem[Kormendy(1977)]{kormendy77} Kormendy, J.\ 1977, \apj, 217, 406
\bibitem[Kormendy et al.(2009)]{kormendy09} Kormendy, J., Fisher, D.~B., Cornell, M.~E. et al.\ 2009, \apjs, 182, 216 
\bibitem[Lambas et al.(1988)]{lambas88} Lambas, D.~G., Groth, E.~J., \& Peebles, P.~J.~E.\ 1988, \aj, 95, 996 
\bibitem[Lauer et al.(2007)]{lauer07} Lauer, T. R., Faber, S. M., Richstone, D. et al.\ 2007, \apj, 662, 808 
\bibitem[Lin \& Mohr(2004)]{lin04} Lin, Y.-T., \& Mohr, J.~J.\ 2004, \apj, 617, 879 
\bibitem[Lin et al.(2009)]{lin09} Lin, Y.-T., Ostriker, J.~P., \& Miller, C.~J.\ 2009, arXiv:0904.3098 
\bibitem[Liu et al.(2008)]{liu08} Liu, F.~S., Xia, X.~Y., Mao, S. et al.\ 2008, \mnras, 385, 23
\bibitem[Liu et al.(2009)]{liu09} Liu, F.~S., Mao, S., Deng, Z.~G. et al.\ 2009, \mnras, 396, 2003 
\bibitem[Loh \& Strauss(2006)]{loh06} Loh, Y.-S., \& Strauss, M.~A.\ 2006, \mnras, 366, 373 
\bibitem[Loubser et al.(2009)]{loubser09} Loubser, S.~I., S{\'a}nchez-Bl{\'a}zquez, P., Sansom, A.~E. et al.\ 2009, \mnras, 398, 133
\bibitem[Mancini et al.(2010)]{mancini10} Mancini, C., Daddi, E., Renzini, A. et al.\ 2010, \mnras, 401, 933 
\bibitem[Matthews, Morgan \& Schmidt(1964)]{matthews64} Matthews, T.~A., Morgan, W.~W., \& Schmidt, M.\ 1964, \apj, 140, 35
\bibitem[McGlynn \& Ostriker(1980)]{mcglynn80} McGlynn, T.~A., \& Ostriker, J.~P.\ 1980, \apj, 241, 915 
\bibitem[M\'endez-Abreu et al.(2008)]{mendezabreu08} M\'endez-Abreu J., Aguerri J.A.L, Corsini E. M et al., 2008, \aap, 478, 353
\bibitem[Merritt(1985)]{merritt85} Merritt, D.\ 1985, \apj, 289, 18 
\bibitem[Merritt et al.(2004)]{merritt04} Merritt, D., Piatek, S., Portegies Zwart, S. et al.\ 2004, \apjl, 608, L25 
\bibitem[Miley et al.(2006)]{miley06} Miley, G.K., Overzier, R.A.; Zirm, A.W., et al.\ 2006, \apjl, 650, L29 
\bibitem[Moore et al.(1996)]{moore96} Moore, B., Katz, N., Lake, G. et al.\ 1996, \nat, 379, 613
\bibitem[Mullis et al.(2003)]{mullis03} Mullis, C. R., McNamara, B. R., Quintana, H., et al.\ 2003, \apj, 594, 154
\bibitem[Murante et al.(2007)]{murante07} Murante, G., Giovalli, M., Gerhard, O. et al..\ 2007, \mnras, 377, 2 
\bibitem[Nelson et al.(2002)]{nelson02} Nelson, A.~E., Simard, L., Zaritsky, D. et al.\ 2002, \apj, 567, 144 
\bibitem[Niederste-Ostholt et al.(2010)]{niederste-ostholt10} Niederste-Ostholt, M., Strauss, M.~A., Dong, F. et al.\ 2010, \mnras, 560
\bibitem[Oemler(1973)]{oemler73} Oemler, A.\ 1973, \apj, 180, 11 
\bibitem[Oemler(1976)]{oemler76} Oemler, A., Jr.\ 1976, \apj, 209, 693 
\bibitem[Ostriker \& Tremaine(1975)]{ostriker75} Ostriker, J.~P., \& Tremaine, S.~D.\ 1975, \apjl, 202, L113 
\bibitem[Ostriker \& Hausman(1977)]{ostriker77} Ostriker, J.~P., \& Hausman, M.~A.\ 1977, \apjl, 217, L125 
\bibitem[Patel et al.(2006)]{patel06} Patel, P., Maddox, S., Pearce, F.~R. et al.\ 2006, \mnras, 370, 85
\bibitem[Poggianti(1997)]{poggianti97} Poggianti, B.~M.\ 1997, \aaps, 122, 399
\bibitem[Ponman et al.(1994)]{ponman94} Ponman, T.~J., Allan, D.~J., Jones, L.~R., et al.\ 1994, \nat, 369, 462 
\bibitem[Popesso et  al.(2006)]{popesso06} Popesso, P., Biviano, A., B{\"o}hringer, H. et al., \ 2006, \aap, 445, 29
\bibitem[Postman \& Lauer(1995)]{postman95} Postman, M., \& Lauer, T.~R.\ 1995, \apj, 440, 28
\bibitem[Reiprich \& Bohringer(2002)]{reiprich02} Reiprich, T.~H., \& Bohringer, H.\ 2002, \apj, 567, 716
\bibitem[Reyes et al.(2008)]{reyes08} Reyes, R., Mandelbaum, R., Hirata, C. et al.\ 2008, \mnras, 390, 1157 
\bibitem[Sandage(1972a)]{sandage72a} Sandage, A.\ 1972a, \apj, 173, 485 
\bibitem[Sandage(1972b)]{sandage72} Sandage, A.\ 1972, \apj, 178, 1 
\bibitem[Sanderson et al.(2009)]{sanderson09} Sanderson, A.~J.~R., Edge, A.~C., \& Smith, G.~P.\ 2009, \mnras, 398, 1698 
\bibitem[Scannapieco \& Tissera(2003)]{scannapieco03} Scannapieco, C., \& Tissera, P.~B.\ 2003, \mnras, 338, 880 
\bibitem[Schombert(1986)]{schombert86} Schombert, J.~M.\ 1986, \apjs, 60, 603 
\bibitem[Schombert(1987)]{schombert87} Schombert, J.~M.\ 1987, \apjs, 64, 643 
\bibitem[Schombert(1988)]{schombert88} Schombert, J.~M.\ 1988, \apj, 328, 475 
\bibitem[Seigar, Graham \& Jerjen(2007)]{seigar07} Seigar, M.~S., Graham, A.~W., \& Jerjen, H.\ 2007, \mnras, 378, 1575 
\bibitem[S\'ersic(1968)]{sersic68} S\'ersic, J.L., 1968, Atlas de Galaxias Australes (C\'ordoba: Obs. Astron. Univ. Nac. C\'ordoba)
\bibitem[Shan et al.(2010)]{shan10} Shan, H., Qin, B., Fort, B. et al.\ 2010, \mnras, 406, 1134 
\bibitem[Silich et al.(2010)]{silich10} Silich, S., Tenorio-Tagle, G., Mu{\~n}oz-Tu{\~n}{\'o}n, C. et al.\ 2010, \apj, 711, 25 
\bibitem[Smith et al.(2005)]{smith05} Smith, G.~P., Kneib, J.-P., Smail, I. et al.\ 2005, \mnras, 359, 417
\bibitem[Smith et al.(2010)]{smith10} Smith, G.~P.; Khosroshahi, H.~G.; Dariush, A., et al.\ 2010, \mnras, in press 
\bibitem[Stanford, Eisenhardt \& Dickinson(1998)]{stanford98} Stanford, S.~A., Eisenhardt, P.~R., \& Dickinson, M.\ 1998, \apj, 492, 461 
\bibitem[Stott et al.(2008)]{stott08} Stott, J.~P., Edge, A.~C., Smith, G.~P. et al.\ 2008, \mnras, 384, 1502
\bibitem[Stott et al.(2010)]{stott10} Stott, J.~P., Collins, C. A., SahlŽn, M., et al.\ 2010, \apj, 718, 23 
\bibitem[Tenorio-Tagle et al.(2005)]{tenorio-tagle05} Tenorio-Tagle, G., Silich, S., Rodr{\'{\i}}guez-Gonz{\'a}lez, A. et al.\ 2005, \apjl, 628, L13 
\bibitem[Tremaine \& Richstone(1977)]{tremaine77} Tremaine, S.~D., \& Richstone, D.~O.\ 1977, \apj, 212, 311 
\bibitem[Trujillo et al.(2001)]{trujillo01} Trujillo, I., Graham, A.~W., \& Caon, N.\ 2001, \mnras, 326, 869 
\bibitem[Trujillo et al.(2006)]{trujillo06} Trujillo, I., F{\"o}rster Schreiber, N.~M., Rudnick, G., et al.\ 2006, \apj, 650, 18 
\bibitem[Trujillo et al.(2007)]{trujillo07} Trujillo, I., Conselice, C.~J., Bundy, K. et al.\ 2007, \mnras, 382, 109 
\bibitem[van Dokkum et al.(1998)]{vandokkum98} van Dokkum, P.~G., Franx, M., Kelson, D.~D. et al.\ 1998, \apj, 500, 714
\bibitem[van Dokkum et al.(2010)]{vandokkum10} van Dokkum, P.G., Whitaker, K.E., Brammer, G., et al.\ 2010, \apj, 709, 1018
\bibitem[Varela et al.(2009)]{varela09} Varela, J., D'Onofrio, M., Marmo, C., et al.\ 2009, \aap, 497, 667
\bibitem[Vikhlinin et al.(2005)]{vikhlinin05} Vikhlinin, A., Markevitch, M., Murray, S.~S. et al.\ 2005, \apj, 628, 655 
\bibitem[Vikhlinin et al.(2006)]{vikhlinin06} Vikhlinin, A., Kravtsov, A., Forman, W. et al.\ 2006, \apj, 640, 691
\bibitem[Vikram et al.(2009)]{vikram09} Vikram, V., Wadadekar, Y., Kembhavi, A.~K. et al.\ 2009, \mnras, L355
\bibitem[Whiley et al.(2008)]{whiley08} Whiley, I. M.; Arag—n-Salamanca, A.; De Lucia, G., et al.\ 2008, \mnras, 387, 1253
\bibitem[Zibetti et al.(2005)]{zibetti05} Zibetti, S., White, S.~D.~M., Schneider, D.~P. et al.\ 2005, \mnras, 358, 949 
\end{thebibliography}
\end{document}